\documentclass[12pt]{article}
\usepackage{epsfig,cite}
\usepackage{amsmath}
\usepackage{amssymb}
\usepackage{color}
\usepackage{graphicx}
\usepackage{verbatim,mathrsfs,subfigure,feynmf}
\usepackage{url}
\usepackage{ytableau}
\include{epsf}
\newcount\timecount
\newcount\hours \newcount\minutes  \newcount\temp \newcount\pmhours
\hours = \time
\divide\hours by 60
\temp = \hours
\multiply\temp by 60
\minutes = \time
\advance\minutes by -\temp
\def\hour{\the\hours}
\def\minute{\ifnum\minutes<10 0\the\minutes
            \else\the\minutes\fi}
\def\clock{
\ifnum\hours=0 12:\minute\ AM
\else\ifnum\hours<12 \hour:\minute\ AM
      \else\ifnum\hours=12 12:\minute\ PM
            \else\ifnum\hours>12
                 \pmhours=\hours
                 \advance\pmhours by -12
                 \the\pmhours:\minute\ PM
                 \fi
            \fi
      \fi
\fi
}

\def\monthname{\relax\ifcase\month 0/\or January\or February\or
   March\or April\or May\or June\or July\or August\or September\or
   October\or November\or December\else\number\month/\fi}

\def\bold#1{\setbox0=\hbox{$#1$}%
     \kern-.025em\copy0\kern-\wd0
     \kern.05em\copy0\kern-\wd0
     \kern-.025em\raise.0433em\box0 }

\def\beq{\begin{equation}}
\def\eeq{\end{equation}}

\def\ga{\mathrel{\raise.3ex\hbox{$>$\kern-.75em\lower1ex\hbox{$\sim$}}}}
\def\la{\mathrel{\raise.3ex\hbox{$<$\kern-.75em\lower1ex\hbox{$\sim$}}}}
\def\gev{{\rm \, Ge\kern-0.125em V}}
\def\tev{{\rm \, Te\kern-0.125em V}}
\def\gyr{{\rm \, G\kern-0.125em yr}}

\def\tbt{\tan \beta}

\def\gappeq{\mathrel{\rlap {\raise.5ex\hbox{$>$}}
{\lower.5ex\hbox{$\sim$}}}}
\def\lappeq{\mathrel{\rlap{\raise.5ex\hbox{$<$}}
{\lower.5ex\hbox{$\sim$}}}}
\def\Toprel#1\over#2{\mathrel{\mathop{#2}\limits^{#1}}}

\def\m12{m_{1\!/2}}

\def\bea{\begin{eqnarray}}
\def\eea{\end{eqnarray}}

\usepackage{lscape}

\def\beq{\begin{equation}}
\def\eeq{\end{equation}}

\def\zthree{\delta_{1i} [g Z_{\chi 2} - g' Z_{\chi 1}]}

\def\zfour{\delta_{2i} [g Z_{\chi 2} - g' Z_{\chi 1}]}

\def\tbt{\tan \beta}

\begin{document}\begin{titlepage}
\pagestyle{empty}
\begin{flushright}

{\tt UMN-TH-4114/22, FTPI-MINN-22/05} \\
\end{flushright}

\begin{center}{\bf \large{Higgsino Dark Matter in Pure Gravity Mediated Supersymmetry}}\\
\vskip 0.2in
{\bf Jason~L.~Evans}$^{a}$ and {\bf Keith~A.~Olive}$^{b}$
\vskip 0.2in
{\small
  {\em $^a$Tsung-Dao Lee Institute, Shanghai Jiao Tong University, Shanghai 200240, China}\\[0.2cm] 
{\em $^b$William I. Fine Theoretical Physics Institute, School of
 Physics and Astronomy,\\ University of Minnesota, Minneapolis, MN 55455,
 USA}}\\

\vspace{0.5cm}
{\bf Abstract}
\end{center}
\baselineskip=18pt \noindent
{\small We consider the prospects for the direct detection of dark matter in pure gravity meditation (PGM) models of supersymmetry breaking. Minimal PGM models
require only two parameters, the gravitino mass, $m_{3/2}$, which sets the UV mass for 
all scalar masses, and $\tbt$. Gaugino masses are generated through anomaly mediation. Typically the lightest supersymmetric state
(the dark matter candidate) is a wino.
Here, we consider a one-parameter extension 
of the minimal model by allowing the Higgs soft masses to deviate from universality.  For simplicity, we take these to be equal and use the $\mu$-term as a surrogate. We also consider non-universal stop masses. When $|\mu| \sim 1$ TeV, the Higgsino is a viable dark matter candidate 
when the gravitino mass is of order $\sim 1$ PeV and $\tbt \simeq 2$. We calculate the spin-dependent and spin-independent cross sections for dark matter scattering on protons. 
For spin-independent scattering, existing 
experimental limits place constraints on the PGM parameter space. Much of the currently allowed parameter space lies above the irreducible neutrino background. Thus, future
direct detection experiments 
will be able to probe much of the remaining PGM parameter space.
}


\vfill
\leftline{February 2022}
\end{titlepage}

\section{Introduction}
\label{sec:intro}
One of the appeals of supersymmetric theories is that they are capable of connecting multiple puzzles of nature within a single framework. For example, supersymmetry explains the radiative stability of the Higgs boson mass \cite{Maiani:1979cx}, has a dark matter candidate \cite{gold,ehnos}, provides for a stable electroweak vacuum \cite{Ellis:2000ig}, is consistent with a 125 GeV Higgs bosons mass \cite{mh}, and leads to rather precise gauge coupling unification and so can explain charge quantization \cite{Ellis:1990zq}. No other model has achieved such a high level of correlation between beyond the standard model puzzles. It truly is a paradigm worthy of our attention.  

However, current constraints coming from the LHC put the scale of supersymmetry breaking masses beyond the TeV mass  \cite{nosusy}. Although this does reduce the protection of the Higgs mass from radiative correction, it does not force us to sacrifice the other successes of supersymmetry. In fact, if we accept some fine tuning in the Higgs sector, some of the challenges of building a successful supersymmetric model can be alleviated, like the Polonyi problem \cite{myy,ego}. This thinking has motivated the study of models like split supersymmetry \cite{split} or pure gravity mediation (PGM) with a large gravitino mass \cite{pgm,pgm2,ArkaniHamed:2012gw,eioy,eioy2,eo,Evans:2014pxa,eno}. In the latter, the only source of supersymmetry breaking in the minimal supersymmetric standard model (MSSM) is the gravitino mass, $m_{3/2}$. Sfermion masses are generated at tree level and so are of order the gravitino mass. The gauginos masses are generated at the one-loop level from anomaly mediation \cite{anom} and so are also proportional to $m_{3/2}$. Even with large sfermion masses in pure gravity mediation, the other salient features of supersymmetric models like a wimp dark matter candidate are preserved since the gauginos are relatively light.

PGM models can be viewed among the most minimal set of supersymmetric models
with radiative electroweak symmetry breaking (EWSB) \cite{ewsb} and full sfermion mass universality \cite{eioy}.
In the context of minimal supergravity,
in addition to the gaugino masses, supersymmetry breaking tri-linear terms, $A_0$, are also set by anomaly mediated interactions and are also proportional to the gravitino mass. Supergravity conditions then fix the Higgs bi-linear supersymmetry breaking term, $B_0 = A_0 - m_{3/2} \simeq -m_{3/2}$ \cite{bfs,vcmssm}. 
Minimization of the Higgs potential then fixes the $\mu$-term and the ratio of Higgs vevs, $\tan \beta$. This single parameter model is however over-constrained, but can be relaxed by adding a Giudice-Masiero (GM) term \cite{gm,ikyy,dmmo} to the K\"ahler potential. In practice, one can then trade the GM term for $\tan \beta$ and use the EWSB conditions to determine the GM coupling. This leaves two free parameters, $m_{3/2}$ and $\tan\beta$. 

The typical and  most widely studied dark matter candidate in pure gravity mediated models is the wino. Because of the anomaly  mediated gaugino mass conditions, the wino tends to be the lightest supersymmetric particle (LSP). Its relic density, determined by annihilations, is relatively insensitive to the sfermion mass spectrum, and a wino of mass of order 3 TeV can lead to the correct relic density \cite{winomass1,winomass,mc13} after inclusion of Sommerfeld enhancement effects~\cite{Sommerfeld1931}.  Furthermore, if the wino is the dark matter candidate and produced from thermal freeze-out, there is a clear prediction for the sfermion mass spectrum. Indeed, the correct relic density ($\Omega h^2 \approx 0.12$ \cite{Planck18}) is obtained in PGM models when $m_{3/2} \sim 800$ TeV \cite{Evans:2014pxa}, making the prospects for the detection of sfermions at the LHC practically non-existent. Instead, we expect relatively strong signals from indirect dark matter experiments. In fact, there is currently some tension with the H.E.S.S. experiment unless the galaxy dark matter profile is of the most cored variety\cite{Cohen:2013ama,Fan:2013faa,Baumgart:2014saa}. This has lead to concerns about the viability of wino dark matter. 

Although these arguments do not decisively rule out wino dark matter, the exploration of other dark matter candidates in models like pure gravity mediation are warranted \cite{Evans:2014pxa}. For example, by including additional vector-like multiplets \cite{eo}, it is possible that either the bino or gluino becomes the LSP.  In this work, we revisit the simpler possibility of Higgsino dark matter in pure gravity mediation \cite{Evans:2014pxa}. 
The Higgsino becomes the LSP when $\mu \ll m_{\tilde W}$. This can be achieved if one \cite{nuhm1} or both \cite{nuhm2}
of the Higgs soft masses is non-universal and differ from $m_{3/2}$ \cite{eioy2}. 
Higgsino dark matter in related models has been discussed in \cite{osi,Delgado:2020url,Dutta:2005zz,Nagata:2014wma,SuryanarayanaMummidi:2020ydm,Co:2021ion}.
Here, we specifically concentrate on the prospects of the direct detection of dark matter in PGM models.  

As we will show, the prospect for the direct detection
of a Higgsino dark matter candidate in PGM models is quite good. This is due to the fact that gauginos are necessarily relatively light as well. This leads to a non-trivial mixing angle between the bino, wino, and Higgsinos. Since LSP-nucleon scattering comes at the cost of a single Higgsino-bino or Higgsino-wino mixing angle, the cross section can be large enough to be measured in direct detection experiments. In fact, much of the parameter space of Higgsino dark matter in PGM will be probed by the upcoming direct detection experiments.  

In what follows, we briefly review the basic idea behind pure gravity mediation in Section
2. While the minimal model can be described by two parameters, $m_{3/2}$ and $\tbt$, the models considered here contain a third parameter, $\mu$ which is equivalent to 
non-universal Higgs masses. Furthermore, we also consider the effect of allowing the stop mass to be non-universal. As in earlier work \cite{Evans:2014pxa}, we find a viable Higgsino-like dark matter candidate with acceptable Higgs masses when $m_{3/2} \sim 1$ PeV, $\tbt \sim 2$, and $|\mu| \sim 1$ TeV. 
In Section 3, we describe our calculation of the scattering cross sections. Our results are presented in Section 4 and our conclusions are given in Section 5.

\section{Pure Gravity Mediation}
In this section, we discuss a few details of minimal pure gravity mediation and its realization with non-universal soft masses. For more details see \cite{eioy2,Evans:2014pxa}.  

The core idea of pure gravity mediation is that all supersymmetry breaking in the visible sector comes from gravity. 
In this scenario, the soft masses are generated the same way they are in minimal supergravity (mSUGRA), ie. with a flat K\"ahler potential, by a correction to the scalar potential coming from Planck suppressed operators, and are of order $m_{3/2}$. In mSUGRA, we expect scalar mass universality at the scale supergravity is broken. This avoids problems with flavor changing neutral currents \cite{en}. In PGM, we take universal scalar masses only for simplicity, since as we will see, the large values needed for $m_{3/2}$ preclude the problems with flavor changing neutral currents.  Since these masses are proportional to the expectation value of the superpotential, they do not a priori depend explicitly on any singlets. The gaugino masses, on the other hand, for models like mSUGRA, do generally depend on a singlet (through a non-trivial gauge kinetic function). In order to get a gaugino mass of order $m_{3/2}$ from gravitational interactions, a singlet supersymmetry breaking field which couples to the gauge fields of the following form is needed,
\begin{eqnarray}
W\supset \frac{c Z}{M_P} {\cal WW}~.
\end{eqnarray}
If the field is not a singlet, this term is forbidden and the leading order contribution is of higher order and thus very suppressed. 

With the gravity mediated contribution to the gauginos masses suppressed, the anomaly mediated contribution \cite{anom} becomes dominant.  In this case, gaugino masses take the form
\begin{eqnarray}
    M_{1} &=&
    \frac{33}{5} \frac{g_{1}^{2}}{16 \pi^{2}}
    m_{3/2}\ ,
    \label{eq:M1} \\
    M_{2} &=&
    \frac{g_{2}^{2}}{16 \pi^{2}} m_{3/2}  \ ,
        \label{eq:M2}     \\
    M_{3} &=&  -3 \frac{g_3^2}{16\pi^2} m_{3/2}\ ,
    \label{eq:M3}
\end{eqnarray}
where $g_i$ are the electroweak and strong gauge couplings and we see that the gaugino masses are loop suppressed compared to sfermion masses. 
Note that contributions from Higgsino loops can be neglected if $\mu \ll m_{3/2}$.

As noted above, in the minimal setup of PGM, the K\"ahler potential is taken to be flat. Scalar masses are universal and equal to gravitino mass at some high energy input scale which we take here to be the scale of gauge coupling unification. Gaugino masses and supersymmetry breaking tri-linear $A$-terms are loop suppressed. Additionally, from what we know in mSUGRA, the $B$-term is fixed at the input scale to be 
$B_0 = A_0 - m_{3/2} \approx -m_{3/2}$. Thus the only free parameters are $m_{3/2}$, $\tan \beta$, and $\mu$.  However, the minimization of the Higgs potential then fixes two of these, leaving a single parameter theory which is overly-restrictive. If a GM term is included in the K\"ahler potential, 
\beq
\Delta K = c_H H_1 H_2  + h.c. \, ,
\label{gmk}
\eeq
where $c_H$ is a constant,
the expressions for $\mu$ and $B$ are modified at the input scale
\begin{eqnarray}
 \mu &=& \mu_0 + c_H m_{3/2}\ ,
 \label{eq:mu0}
 \\
  B\mu &=&  \mu_0 (A_0 - m_{3/2}) + 2 c_H m_{3/2}^2\ .
   \label{eq:Bmu0}
\end{eqnarray}
As a result, 
after minimization of the Higgs potential,  there are two remaining free parameters\footnote{See \cite{eioy} for more details.} which can be chosen to be $\tan\beta$ and $m_{3/2}$
(so that $\mu$ and $c_H$ are determined by EWSB). 
Generically, this minimal model has a wino dark matter candidate\footnote{Even in this minimal setup, a Higgsino dark matter candidate can be realized in a focus point-like scenario \cite{fp}. However, as the fully universal mSUGRA-PGM model is a special case of our consideration here, it is included in our analysis.} with the correct thermal relic density for a gravitino mass of order 800 TeV.

In this work, we examine a slightly more generic scenario with non-universal Higgs masses \cite{eioy2,Evans:2014pxa}. This can be accomplished by the addition of higher dimensional operators involving the supersymmetry breaking field and the Higgs fields. We take the following K\"ahler potential 
\begin{eqnarray}
K =   y y^*  + K^{(H)} +  K^{(Z)}  +  \log |W|^2~,
\label{K1}
\end{eqnarray}
where
\begin{eqnarray}
K^{(Z)} = Z Z^* (1-\frac{Z Z^*}{\Lambda^2})~,
\label{kz}
\end{eqnarray}
and
\begin{eqnarray}
K^{(H)} = (1 + a \frac{Z Z^*}{M_P^2}) H_1 H_1^*
+ (1 + b \frac{Z Z^*}{M_P^2}) H_2 H_2^* + (c_H H_1 H_2 + h.c.) \, ,
\label{kh}
\end{eqnarray}
and $y$ represents all other MSSM fields other than the Higgs bosons, $H_1,H_2$, $W$ is the MSSM superpotential, and $Z$ is the supersymmetry breaking field which we assume is strongly stabilized at some mass scale $\Lambda < M_P$ \cite{dine,dlmmo,ego}. We do not contemplate the origin of these higher dimensional operators and consider this only as an effective theory. For this K\"ahler potential, all MSSM sfermion fields gets soft masses equal to $m_{3/2}$ with the exception of the Higgs soft masses. If $Z$ is a Polonyi-like field, its expectation value  induces an $F$-term and both the Higgs soft masses become free parameters given by the constants $a$ and $b$.  The two soft Higgs masses are
$m_{H_1}^2 = (1-3a)m_{3/2}^2$ and $m_{H_2}^2 = (1-3b)m_{3/2}^2$.

Our ability to obtain a Higgsino dark matter candidate is simplified if we allow $\mu$ to be a free input parameter. Thus we require only one
additional input parameter and we take $m_{H_1}^2 =m_{H_2}^2$ or $a = b$.
Typically, the Higgs masses are set at the GUT scale and then are renormalization group evolved to the weak scale. The Higgs potential minimization conditions are then used to determine $\mu$ and in this case $c_H$ (since $B_0$ is fixed in mSUGRA and we keep $\tan \beta$ free). We will instead use $\mu$ as an input and use the Higgs potential minimization conditions to determine $m_{H_2}^2$ and $c_H$ and $m_{H_1}^2$ is set equal to $m_{H_2}^2$ at the GUT scale. This implementation of non-universal Higgs masses has the advantage of being able to readily realize a Higgsino mass which gives the correct thermal relic density. 

Later we will consider a slightly more generic model where the stop soft masses are equal but different from the Higgs soft masses and other sfermion masses. This can be accomplished in the effective theory in a similar manner as was done in Eq. (\ref{kh}) for the Higgs masses. The purpose of considering non-universal stop masses is to attempt to capture the features of more generic models of pure gravity mediation beyond the minimal model. In pure gravity mediation models, because the soft masses are so large, order one changes in the soft masses have minimal effect on the low-scale theory except for the stop masses and Higgs masses. The Higgs boson masses play an important role in radiative electroweak symmetry breaking and in determining the mass of the Higgsino.  The stop masses, on the other hand, play an important role in determining the standard model Higgs boson mass. By taking the Higgs and stop soft masses non-universal, we are able to capture the major features of a more generic pure gravity mediation model and see how Higgsino dark matter would look in this scenario. Thus, for this work we take a model with the following set of free parameters
\begin{eqnarray}
m_{3/2}, \quad \quad \tan\beta, \quad \quad \mu, \quad \quad m_T~,
\end{eqnarray}
where the gravitino mass is the mass of all the scalar particles at GUT scale except the stops and the Higgs bosons,  and $m_T$ is the mass of left- and right-handed stop masses at the input scale (thus also equal to $m_{{\tilde b}_L}$, while $m_{{\tilde b}_R} = m_{3/2}$). Initially, we will take $m_T=m_{3/2}$ and then later consider this full set of parameters. 

\begin{figure}[!ht]
\centering
\includegraphics[width=8cm]{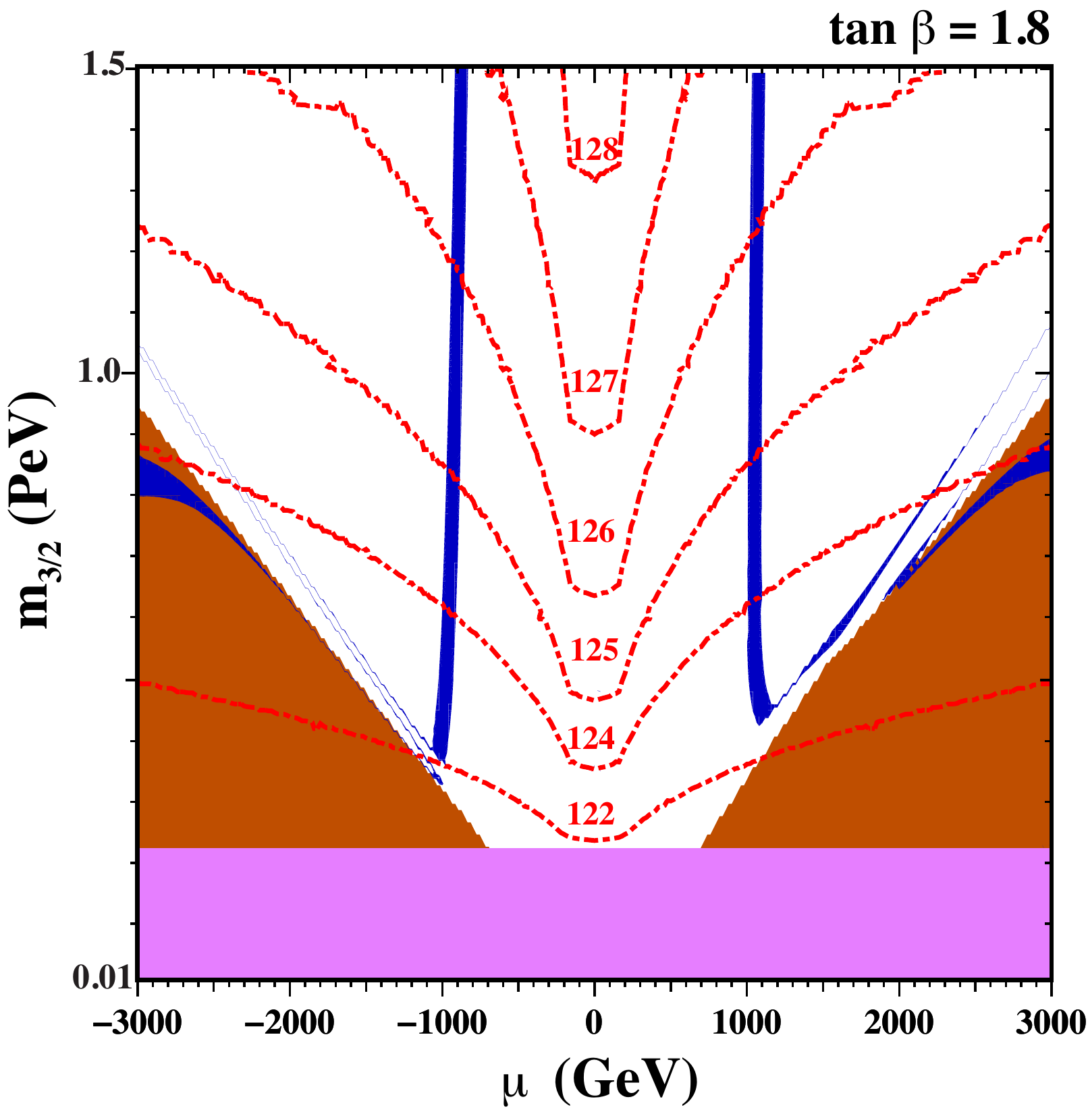}
\includegraphics[width=8cm]{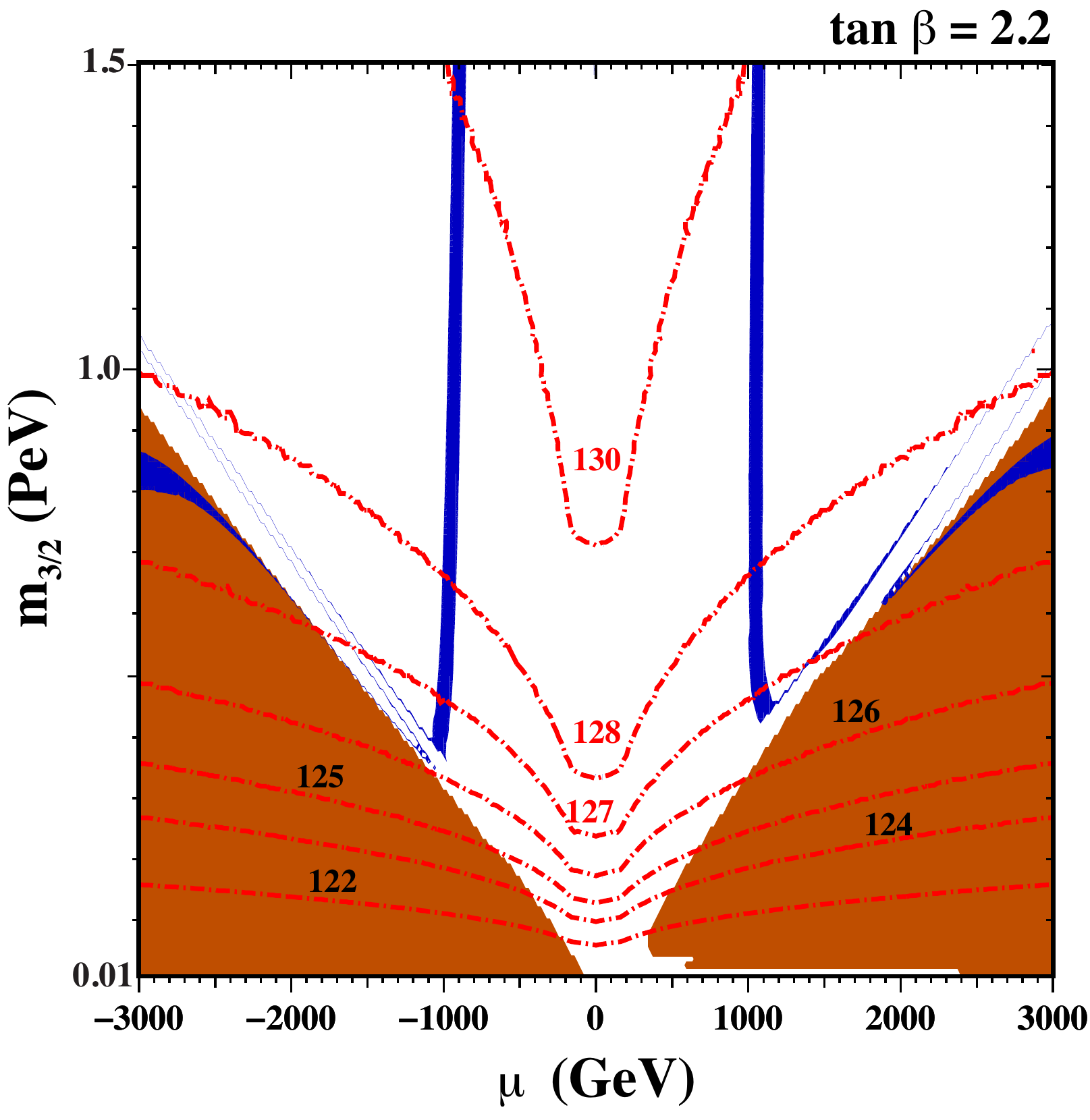}
\caption{\it The $(\mu, m_{3/2})$ plane for fixed $\tan \beta = 1.8$ (left) and 2.2 (right). The pink shaded region is excluded
as it contains a tachyonic stop. In the dark red shaded region, there is a wino LSP. In the remainder of the plane, the Higgsino is the LSP. Higgs mass contours, with masses labelled are shown as red dot-dashed curves. In the dark blue strips, the relic density is $\Omega_\chi h^2 = 0.11-0.13$. 
}
\label{oldplanes}
\end{figure}

The restricted parameter space (with $m_T = m_{3/2}$) was previously studied \cite{Evans:2014pxa} and we present some updated results in the $(\mu,m_{3/2})$ planes shown in Fig.~\ref{oldplanes} for fixed values of $\tan \beta = 1.8$ and 2.2.  For $\tbt = 1.8$ and $m_{3/2} \la 200$ TeV, one of the stops is tachyonic and that excluded region is shaded pink.  The dark red shaded regions contain a wino LSP. In the unshaded regions, there is a Higgsino LSP. The LSP mass contours are not shown but the wino LSP mass is determined primarily from anomaly mediation with $m_{\tilde W} \approx 0.27 m_{3/2}$. The mass of the Higgsino depends primarily on $\mu$, $m_{\tilde H} \approx 1.1 \mu$. For both states, there is a slight dependence on the sign of $\mu$ and $\tbt$. The values of the Higgs mass is shown by a series of red dot-dashed contours as labeled. As one can see, there is a strong dependence of the Higgs mass on $\tbt$. For $\tbt > 2.2$ and the large values of $m_{3/2}$ we considered, the Higgs mass begins to exceed the experimental value even if a relatively
large theoretical uncertainty is assigned to the Higgs mass calculation.  At $\tbt < 1.8$, renormalization group evolution inevitably leads to a tachyonic stop for small $m_{3/2}$ and is excluded.

In the dark blue shaded strips of Fig.~\ref{oldplanes},
the LSP has a relic density $\Omega h^2 = 0.11 - 0.13$.
Note that this range is taken to be significantly wide than the Planck results \cite{Planck18} which indicate a cold dark matter density of $\Omega_c h^2 = 0.1200 \pm 0.0012$ and enhances the visibility of these strips. 
The thermal relic density in the Higgsino region depends primarily on the mass of the Higgsino \cite{osi} and the desired abundance
is achieved when $\mu \approx -900$ GeV and 1080 GeV, with a Higgsino mass
just over 1100 GeV for both positive and negative $\mu$ (the sign dependence on $\mu$ in the Higgsino mass calculation
is due to one-loop threshold corrections).  The preferred regions are seen as the
vertical strips in both panels. In the dark red shaded region at large $|\mu|$ and $m_{3/2} \approx 800$ TeV, the wino is the LSP and
$\Omega h^2 \approx 0.12$.
One also finds three very thin dark blue diagonal strips where the desired relic density can be obtained through coannihilations between the wino and two Higgsinos which are all nearly degenerate in that region.
These have been discussed in more detail in \cite{Evans:2014pxa}.
Our primary interest here are the nearly vertical relic density strips. 

\begin{figure}[!ht]
\centering
\includegraphics[width=8cm]{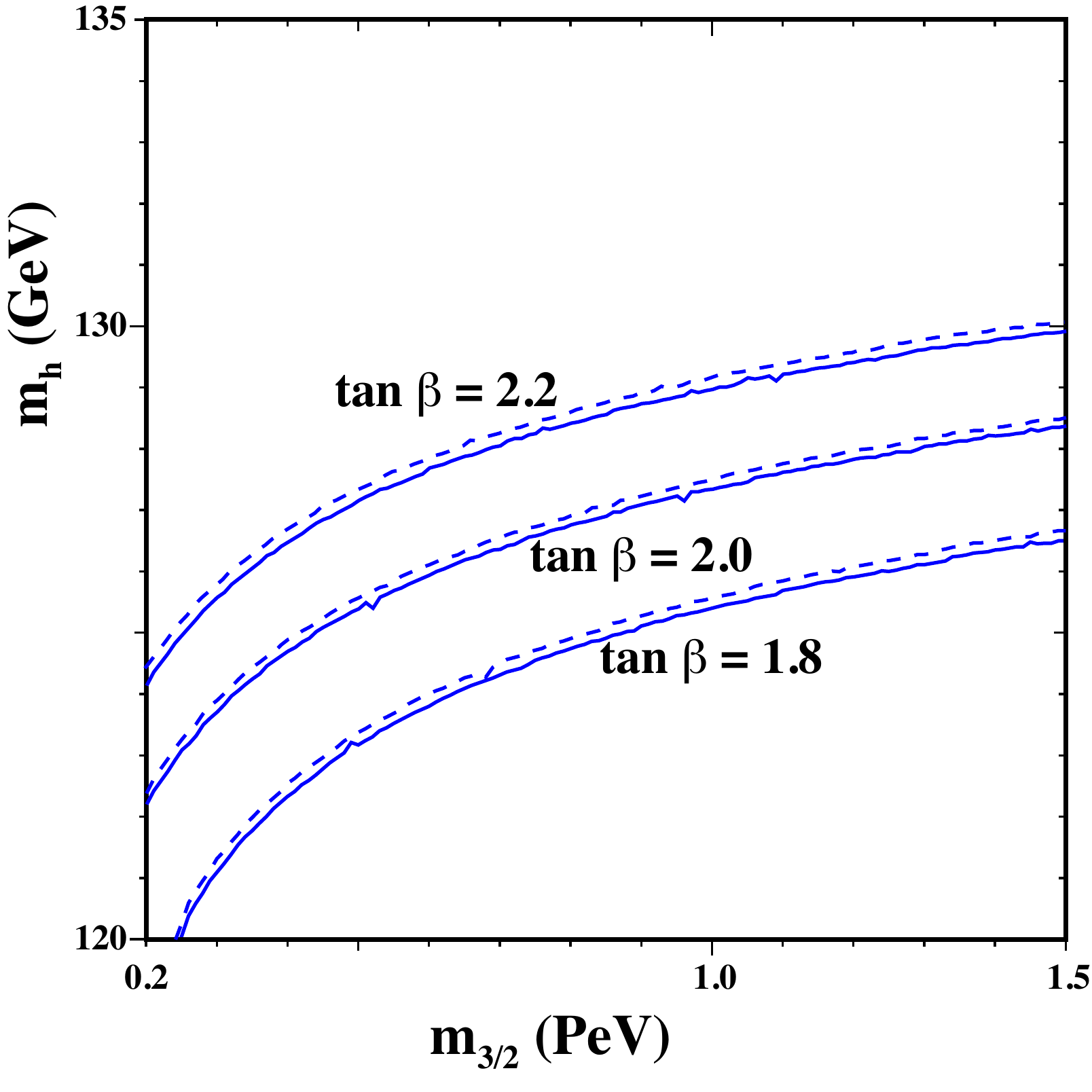}
\includegraphics[width=8cm]{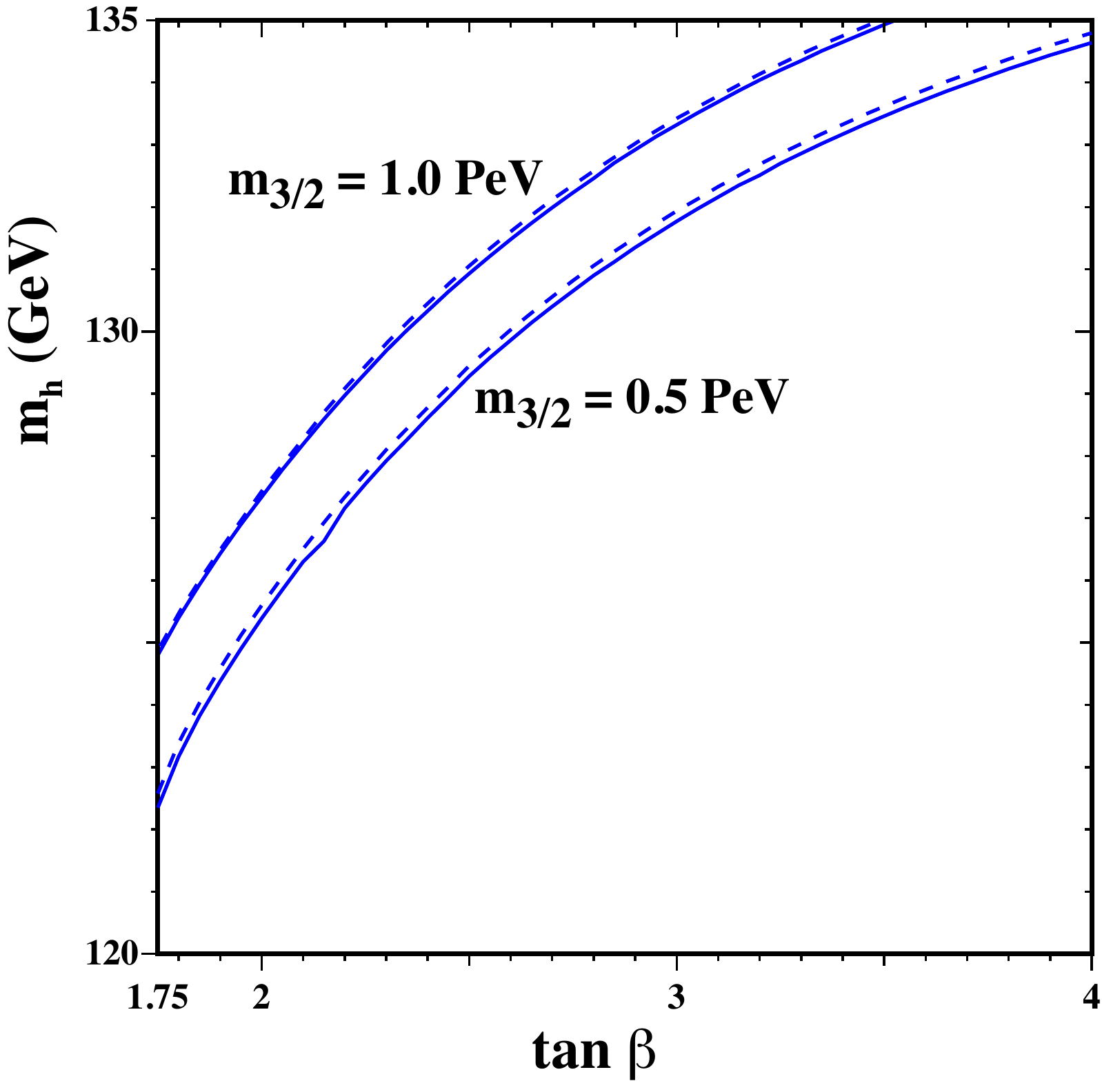}
\caption{\it (left) The Higgs mass vs. $m_{3/2}$ for three values of $\tan \beta = 1.8$, 2.0, and 2.2. (Right)  The Higgs mass vs. $\tbt$ for two values of $m_{3/2} = 0.5$ and 1.0 PeV. The value of $\mu$ is fixed to 1080 GeV (solid) and -900 GeV (dashed). }
\label{mh}
\end{figure}

We show the dependence of the Higgs mass on both the gravitino mass and $\tbt$ in Fig.~\ref{mh}. We plot $m_h$ vs $m_{3/2}$ for three values of $\tbt = 1.8$, 2.0, and 2.2.
The value of $\mu$ is fixed to 1080 GeV and shown as solid curves and -900 GeV as dashed curves.
As one can see, this range of $\tbt$ covers the experimentally acceptable values of $m_h$ for a wide range of gravitino masses.  Note again that the stop mass becomes tachyonic for $\tbt \lesssim 1.8$. In the right panel we see the rapid increase in $m_h$ with $\tbt$, for fixed values of $m_{3/2} = 0.5$ and 1 PeV.

As noted above, the models considered have non-universal Higgs masses, which are set equal to each other at the GUT scale. The degree of non-universality required is shown in Fig.~{\ref{nuhm} which shows the value of 
$a = \frac13 (1- m_{H_1}^2/m_{3/2}^2)$ as a function of the gravitino mass (assuming $m_{H_1} = m_{H_2}$). For the displayed curves, $\mu = 1080$
GeV, and as a function of $m_{3/2}$ they track the right blue shaded region in each panel of Fig.~\ref{oldplanes}. The curves for the left blue shaded regions (with $\mu = -900$ GeV) would be nearly identical. As one can see, for $\tbt = 1.8$, 
universality is only achieved for very large 
$m_{3/2}$ where the Higgs mass is significantly larger than 125 GeV.  However, for $\tbt = 2.2$
complete scalar mass universality is achieved when $m_{3/2} = 0.49$ PeV, and the Higgs mass is 127.1 GeV. We also show the corresponding curve with $\tbt = 2.0$. In this case, universality occurs at $m_{3/2} = 1.35$ PeV, 
but there $m_h = 128.1$ GeV.

\begin{figure}[ht!]
\centering
\includegraphics[width=8cm]{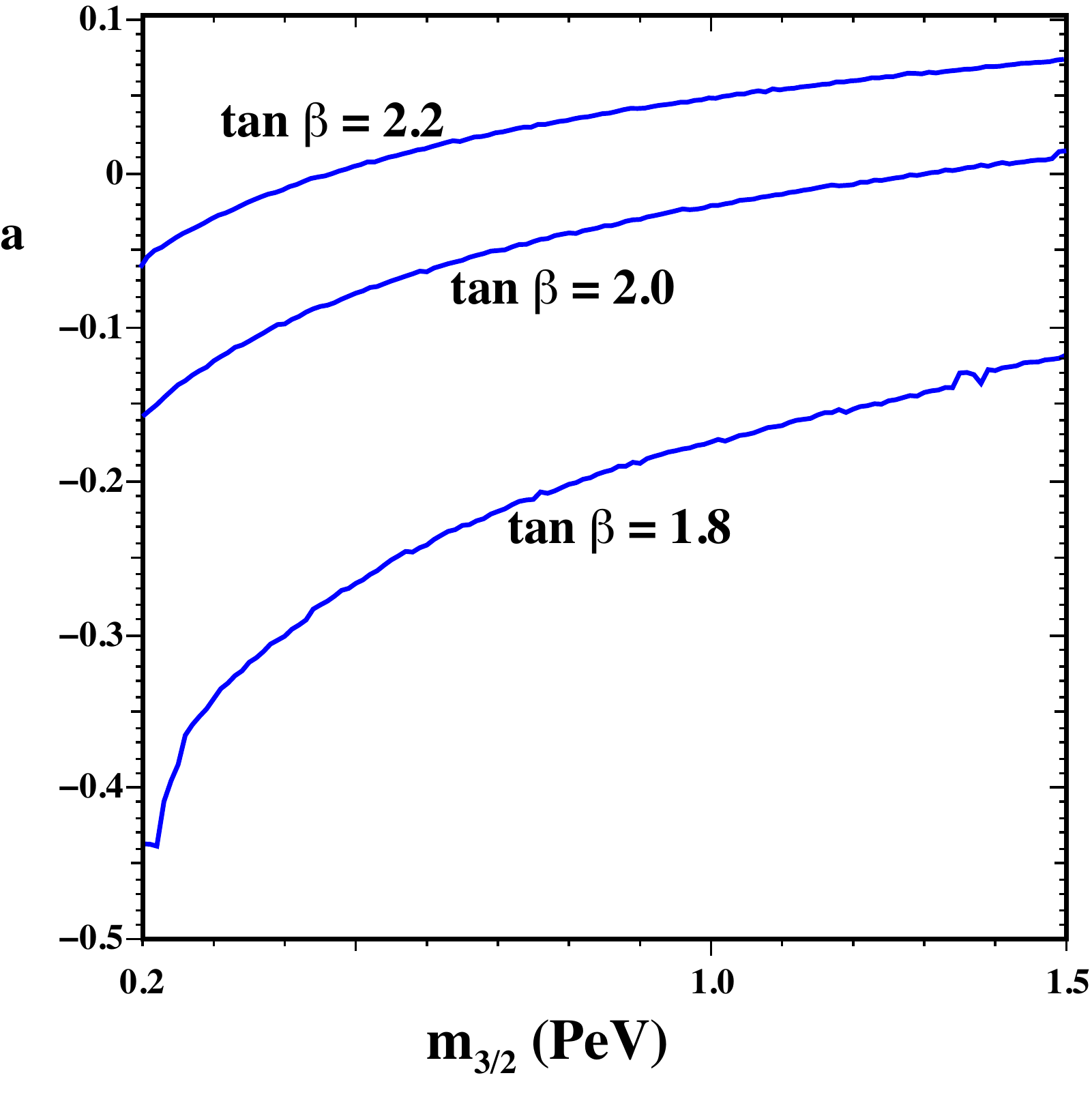}
\includegraphics[width=8cm]{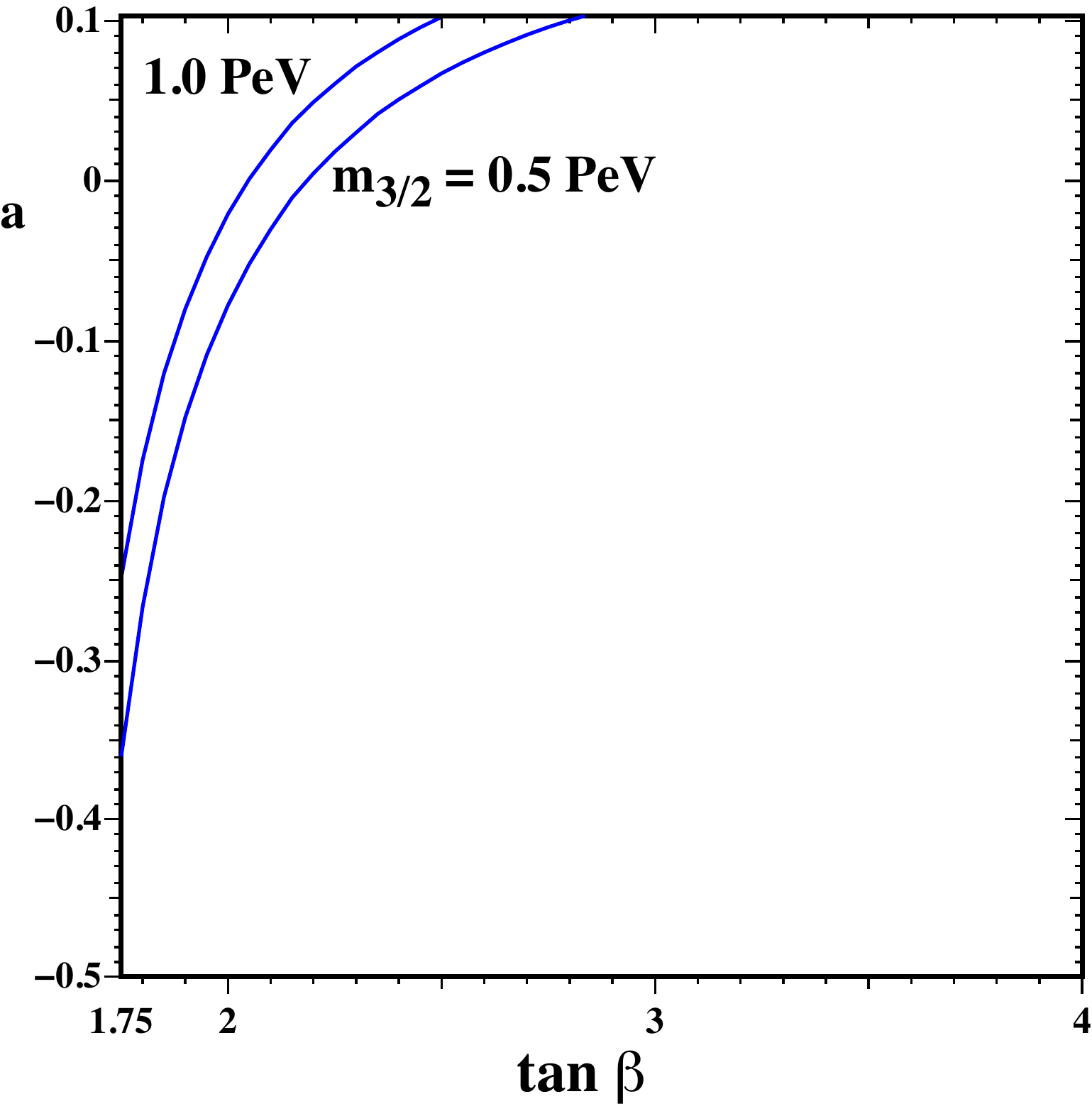}
\caption{\it (left) The degree Higgs mass non-universality for three values of $\tan \beta = 1.8$, 2.0, and 2.2. Plotted is the K\"ahler coupling $a = \frac13 (1 - m_{H_1}^2/m_{3/2}^2) $ vs. $m_{3/2}$.  Full scalar mass universality corresponds to $a = 0$. 
(right) K\"ahler coupling $a$ vs. $\tbt$ for two values of $m_{3/2} = 0.5$ and 1.0 PeV.  This plot assumes $\mu = 1080$ GeV, but would be nearly identical for $\mu = -900$ GeV.   }
\label{nuhm}
\end{figure}

\section{Direct Detection of Higgsino Dark Matter}

It is straight-forward to calculate the elastic scattering cross section
for a neutralino on a nucleon \cite{EF,etal,eflo}. Here we simply quote
the most important ingredients from \cite{eflo} for the purpose of studying Higgsino elastic scattering. We consider only spin-dependent and spin-independent
interaction from
the four-fermi
Lagrangian for $\chi$-nucleon
scattering:
\begin{equation}
{\cal L} \ni \alpha_{2i} \bar{\chi} \gamma^\mu \gamma^5 \chi \bar{q_{i}} 
\gamma_{\mu} \gamma^{5} q_{i} +
\alpha_{3i} \bar{\chi} \chi \bar{q_{i}} q_{i}
\label{lagr}
\end{equation}

If we ignore the contributions from squark exchange 
as well as the heavy Higgs scalar exchange\footnote{Though these contributions are negligible for the cases under study, they are included in all our numerical work.},
the coefficients $\alpha_i$ can be written as
\begin{eqnarray}
\alpha_{2i} & \simeq &  - \frac{g^{2}}{4 m_{Z}^{2} \cos^{2}{\theta_{W}}} \left[
\left| Z_{\chi_{3}} \right|^{2} - \left| Z_{\chi_{4}} \right|^{2}
\right] \frac{T_{3i}}{2}~,
\label{alpha2}
\end{eqnarray}
and
\begin{eqnarray}
\alpha_{3i} & \simeq &  - \frac{g m_{qi}}{4 m_{W} m^{2}_h B_{i}} \left[ Re \left( 
\zthree \right) D_{i} C_{i} \right. \nonumber \\
& & \mbox{} +  Re \left. \left( \zfour \right) 
{D_{i}^{2}}~,
 \right]
\label{alpha3}
\end{eqnarray}
where $T_{3i}$ denotes isospin for 
up type ($i=1$) and down type ($i=2)$ quarks, and $Z_{\chi j}$ corresponds to the ${\tilde B}, {\tilde W}, {\tilde H_1}, {\tilde H_2}$ component of the LSP for $j = 1, 2, 3, 4$ respectively,  and
\beq
\delta_{1i} = Z_{\chi 3} (Z_{\chi 4}) \qquad , \qquad \delta_{2i} = Z_{\chi 4}
(-Z_{\chi 3}),
\label{moredefs1}
\eeq
\beq
B_{i} = \sin{\beta} (\cos{\beta}) \qquad , \qquad 
C_{i} = \sin{\alpha} (\cos{\alpha})\qquad , \qquad  D_{i} = \cos{\alpha} ( -
\sin{\alpha}) ~,
\label{moredefs2}
\eeq
for up (down) type quarks. We denote by $m_h$
the mass of the light scalar Higgs and $ \alpha $ denotes the Higgs mixing
angle.

The mixing angles of a Higgsino LSP with the different neutralino components can be approximated as 
\begin{eqnarray}
&Z_{\chi 1}= -\frac{\sqrt{2}}{2}\left(c_\beta\pm s_\beta\right)\frac{M_Zs_W}{M_1-|\mu|}+{\cal O}(M_Z^3)~,\\
&Z_{\chi 2}=\frac{\sqrt{2}}{2}\left(c_\beta\pm s_\beta\right)\frac{M_Zc_W}{M_2-|\mu|}+{\cal O}(M_Z^3)~,\\ 
&Z_{\chi_3}=\mp \frac{\sqrt{2}}{2} \pm \frac{\sqrt{2}}{8}\left(\frac{c_W^2\left(c_\beta^2-s_\beta^2\right)}{|\mu|(M_2-|\mu|)}+\frac{c_W^2\left(c_\beta\pm s_\beta\right)^2}{(M_2-|\mu|)^2} +\frac{s_W^2\left(c_\beta^2-s_\beta^2\right)}{|\mu|(M_1-|\mu|)}+\frac{s_W^2\left(c_\beta\pm s_\beta\right)^2}{(M_1-|\mu|)^2}\right)M_Z^2+  {\cal O}(M_Z^3)~,\\
&Z_{\chi_4}=~~\frac{\sqrt{2}}{2} + \frac{\sqrt{2}}{8}\left(\frac{c_W^2\left(c_\beta^2-s_\beta^2\right)}{|\mu|(M_2-\mu)}-\frac{c_W^2\left(c_\beta\pm s_\beta\right)^2}{(M_2-\mu)^2} + \frac{s_W^2\left(c_\beta^2-s_\beta^2\right)}{|\mu|(M_1-\mu)}-\frac{s_W^2\left(c_\beta\pm s_\beta\right)^2}{(M_1-\mu)^2}\right)M_Z^2+{\cal O}(M_Z^3)~,
\end{eqnarray}
where $s_\beta=\sin\beta$, $c_\beta=\cos\beta$, $c_W=cos\theta_W$, and $s_W=\sin\theta_W$ and the upper (lower) sign is for $\mu>0$ $(\mu<0)$. This approximation is valid as long as the LSP is predominantly Higgsino and $M_Z/(M_2-|\mu|)\ll 1$. Using these approximate relations and taking the decoupling limit where $\alpha=\beta-\frac{\pi}{2}$, $\alpha_{3i}$ and $\alpha_{2i}$ can be simplified to
\begin{eqnarray}
\alpha_{2i}\simeq  - \frac{1}{8}\cos^2\beta\left(1-\tan^2\beta\right)\left(\frac{g'^2}{|\mu|(M_1-|\mu|)}+\frac{g^2}{|\mu|(M_2-|\mu|)}\right)\frac{T_{3i}}{2}~, \label{eq:al2App}
\end{eqnarray}
and 
\begin{eqnarray}
\alpha_{3i}\simeq  -\frac{1}{8}\frac{m_{q_i}\cos^2\beta\left(1\pm \tan\beta\right)^2 }{m_{H_2}^2}\left(\frac{g'^2}{M_1-|\mu|}+\frac{g^2}{M_2-|\mu|}\right)~. \label{eq:al3App}
\end{eqnarray}
We have checked that numerically these expressions are quite accurate for much of the parameter space we consider. In fact, even for $M_2-|\mu|\sim 400$ GeV, these expressions are still accurate to within about 5$\%$

The spin-independent scattering cross section for the LSP on a proton
can be written as
\begin{equation}
\sigma_{3} = \frac{4 m_{r}^{2}}{\pi}  f_{p}^{2}~,
\label{si}
\end{equation}
where $m_r$ is the reduced LSP mass, and
\begin{equation}
\frac{f_{p}}{m_{p}} = \sum_{q} f_{Tq}^{(p)} 
\frac{\alpha_{3q}}{m_{q}} ~.
\label{f}
\end{equation}
The parameters
$f_{Tq}^{(p)}$  are defined
by
\begin{equation}
m_p f_{Tq}^{(p)} \equiv \langle p | m_{q} \bar{q} q | p \rangle~,
\label{defbq}
\eeq
and have recently been re-evaluated \cite{sospin}
\begin{eqnarray}
f_{Tu}^{(p)} = 0.018 \pm 0.005, \qquad f_{Td}^{(p)} = 0.027 \pm 0.007,
\qquad f_{Ts}^{(p)} = 0.037 \pm 0.017 \nonumber \\
f_{Tc}^{(p)} = 0.078 \pm 0.002, \qquad f_{Tb}^{(p)} = 0.072 \pm 0.002,
\qquad f_{Tt}^{(p)} = 0.069 \pm 0.001
\label{pinput}
\end{eqnarray}

The spin-dependent elastic $\chi$-proton cross section can be
written as
\begin{equation}
\sigma_{2} = \frac{24}{\pi} G_{F}^{2} m_{r}^{2} a_p^2 ~,
\label{sd}
\end{equation}
where
\begin{equation}
a_{p} = \sum_{i} \frac{\alpha_{2i}}{\sqrt{2} G_{f}} \Delta_{i}^{(p)} \, .
\label{a}
\end{equation}
The factors $\Delta_{i}^{(p,n)}$ parametrize the quark spin content of the
nucleon \cite{sospin}
\beq
\Delta_{u}^{(p)} = 0.84 \pm 0.03, \qquad \Delta_{d}^{(p)} = -0.43 \pm
0.03,
\qquad \Delta_{s}^{(p)} = - 0.09 \pm 0.03~.
\label{spincontent}
\eeq

\section{Results}

Despite the very large scalar masses associated with PGM models, the Higgsino 
as a dark matter candidate still lends itself to the possibility of detection 
in direct detection experiments.  As we will see, although some of the parameter space
described above is already excluded by existing data, a substantial range in
parameters remains viable and potentially detectable in future experiments,
as much of the parameter space predicts scattering cross sections in excess of the 
so-called neutrino floor \cite{floor}.

In Fig.~\ref{fig:planeSI}, we show the same ($\mu, m_{3/2}$) planes as in Fig.~\ref{oldplanes} with $\tbt = 1.8$ (2.2) in the left (right) panels,
and provide contours of the spin-independent cross sections, $\sigma_p^{SI} = \sigma_3$.
Contour labels correspond to units of $10^{-8}$ pb. Unlabeled contours correspond to $2 \times$ and $5 \times$ within each decade. We also show the current bound from the PandaX-4T commissioning run \cite{PandaX} as black contours. In the Higgsino region (unshaded),
the allowed parameter space is above the black line, while in the wino region (shaded), it is below the line. These bounds are relevant only
for $\mu >0$ as there are significant cancellations for $\mu < 0$ which suppress the cross section \cite{eflo,ads,Baer,blind}, sometimes known as blind spots. This cancellation can be understood from examining Eq. (\ref{eq:al3App}). Since we consider $\tan\beta$ close to one, $\alpha_{3i}$ is strongly suppressed for $\mu<0$ as compared to $\mu>0$, as can be see in Eq. (\ref{eq:al3App}).
For $\mu \gtrsim 1.1$ TeV, (i.e. to the right of the nearly vertical blue strip), the 
relic Higgsino density from thermal freeze-out exceeds the observed cold dark density
and is only viable if there is an additional (post freeze-out) source of entropy production. In this regime, the bound on the cross section (at roughly $10^{-9}$ pb) weakens slightly as the Higgsino mass ($\mu$) increases. At lower $\mu$, the thermal relic density
is too low and we have scaled the cross section bound by $\Omega_\chi h^2/0.12$ to reflect the lower density (and hence scattering rates). In the event that there is 
a non-thermal source of Higgsinos, the unscaled limit is shown as a dashed black curve
which strengthens significantly as the mass is lowered (either lower $\mu$, or lower $m_{3/2}$ in the wino region). 

\begin{figure}[!ht]
\centering
\includegraphics[width=8cm]{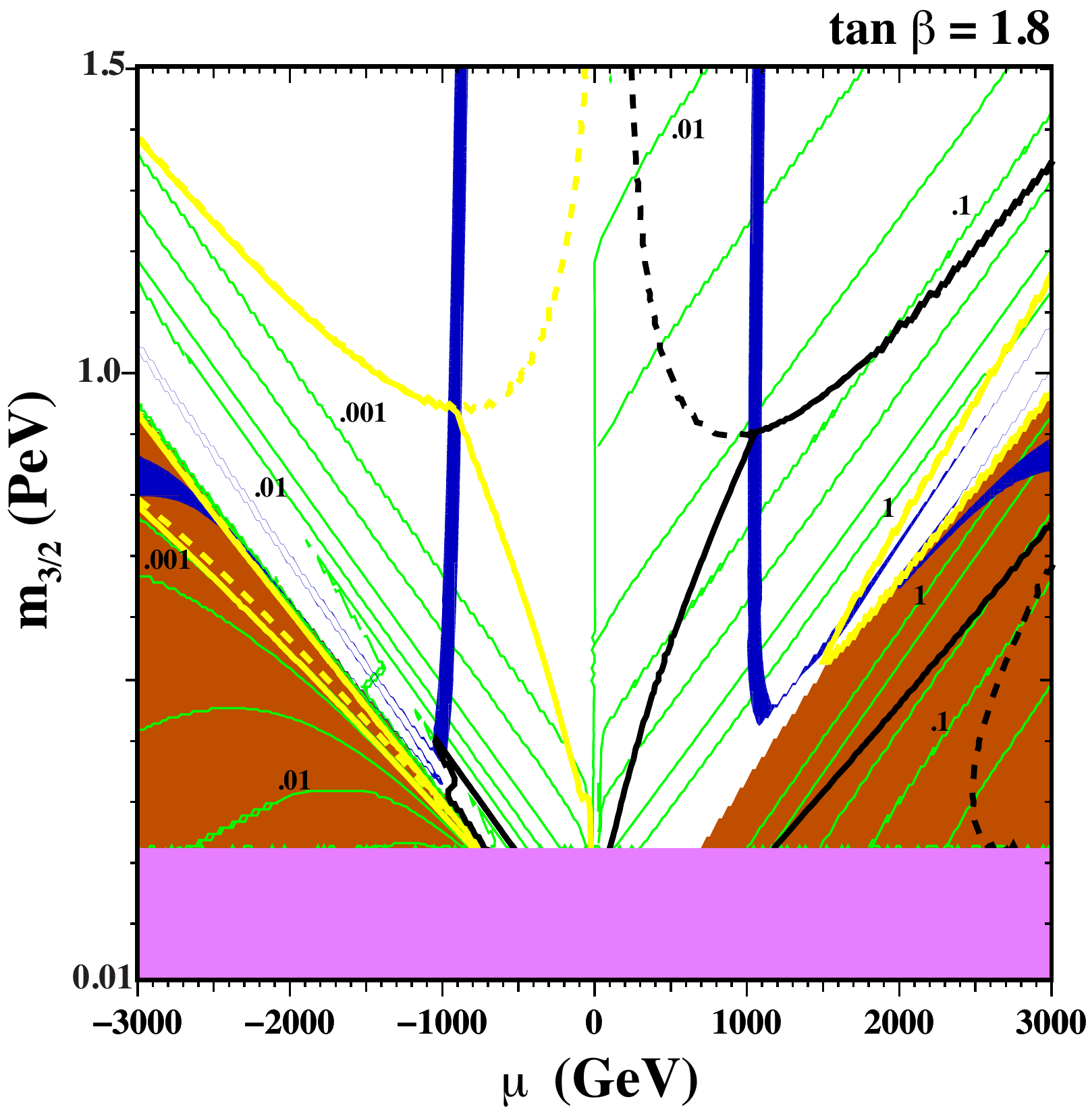}
\includegraphics[width=8cm]{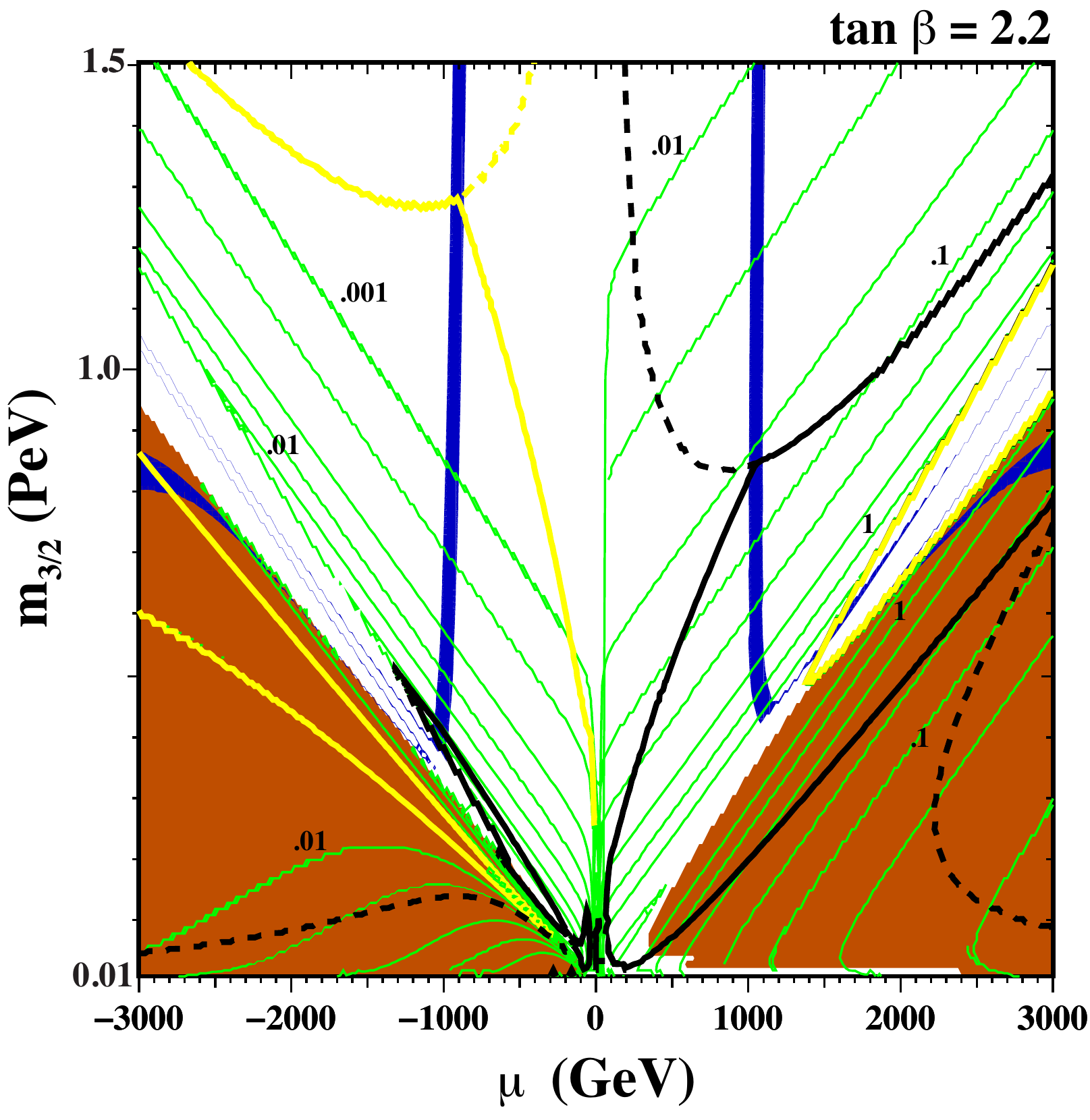}
\caption{\it The spin-independent elastic cross section for LSP scattering on protons in the ($\mu, m_{3/2}$) plane for fixed $\tbt = 1.8$ (left) and 2.2 (right). The shaded regions are the same as in Fig.~\ref{oldplanes}.  Values of the cross sections are as labelled in units of $10^{-8}$ pb. Contours between the labeled decades are 2 and 5 $\times$ the preceding decade. Also shown as thick black curves are contours for the current experimental bound from PandaX-4T \cite{PandaX}. Solid contours correspond to the scaled limit when 
the relic density is $< 0.12$. The neutrino floor is shown by the thick yellow contour.
  }
\label{fig:planeSI}
\end{figure}

Although there is currently no significant experimental constraint on the parameter space
when $\mu < 0$ (except for the small loop around (-1, 300) TeV), much of the parameter space is potentially detectable in future experiments. The yellow contours in Fig. \ref{fig:planeSI} show the position of neutrino floor \cite{floor} below which direct detection becomes overwhelmed by the inevitable neutrino background. For $\mu > 0$, the cross section drops below the neutrino floor only in the wedge near the Higgsino/wino boundary. There is actually a black contour below the yellow as the cross section drops precipitously when the LSP
changes from an anti-symmetric Higgsino ($[{\tilde H}_1 - {\tilde H}_2]/\sqrt{2}$) to a symmetric combination ($[{\tilde H}_1 + {\tilde H}_2]/\sqrt{2}$) \cite{osi} causing a strong cancellation. A similar wedge (without a black contour) is seen when $\mu < 0$ when the LSP changes from a Higgsino to a predominantly wino. The dashed contour again ignores the fact that the relic density is low and does
not include a scaling of the cross section. For $\mu < 0$, outside the wedge,
only portion of the plane above the yellow contour is potentially unobservable. 

Some of this behavior is more easily understood in the one-dimensional plots 
shown in Fig.~\ref{fig:1DSI}. In the left panel, we show the spin-independent cross section as a function of the gravitino mass, while in the right panel as a function of $\tbt$. We have chosen values of $\mu$ so that we obtain the correct thermal relic density
in the Higgsino region, $\mu = 1080$ GeV
(solid curves) and $\mu = -900$ (dashed curves). Consider for example the spin-independent cross section $\mu > 0$.
For low $m_{3/2}$, the wino is the LSP,
and the cross section rises with $m_{3/2}$ to a maximum, and then drops once the LSP 
becomes Higgsino-like. There is little $\tbt$ dependence, which is also seen in the right panel where the curves (for fixed $m_{3/2}$) are rather flat. In both panels, the upper 
horizontal red line represents the current 90 \% upper limit from PandaX-4T \cite{PandaX} for a $\sim 1$ TeV LSP. This limit excluded values of $m_{3/2} \lesssim 850$ TeV for $\mu > 0$. 
The lower horizontal red line corresponds to the neutrino floor for a $\sim 1$ TeV LSP.
All of the predicted cross sections shown lie above the floor and are in principle detectable.  For $\mu < 0$, there are significant cancellations which occur in Eq.~(\ref{alpha3}) which suppress the cross section. The strong dependence on $m_{3/2}$ 
for low mass occurs as the identity of the LSP changes from wino to Higgsino as $m_{3/2}$ is increased. In addition, the LSP changes between ${\tilde H}_{[1,2]}$ and ${\tilde H}_{(1,2)}$ involving rapid changes in the mixing angles
and a strong variation in the cross section. At large values of $m_{3/2}$, the cross section falls beneath the neutrino floor for $\mu < 0$. 

\begin{figure}[!ht]
\centering
\includegraphics[width=8cm]{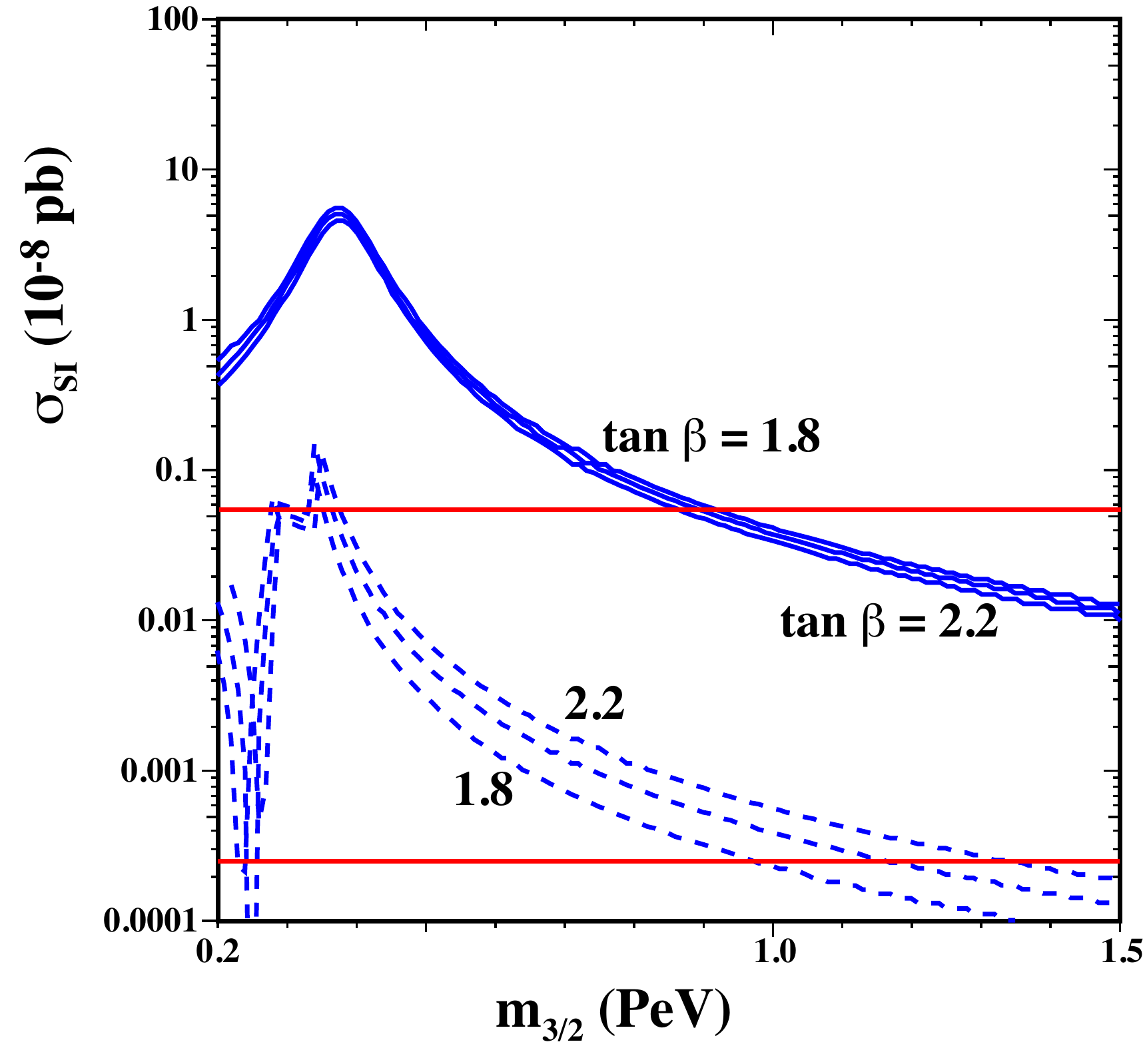}
\includegraphics[width=8cm]{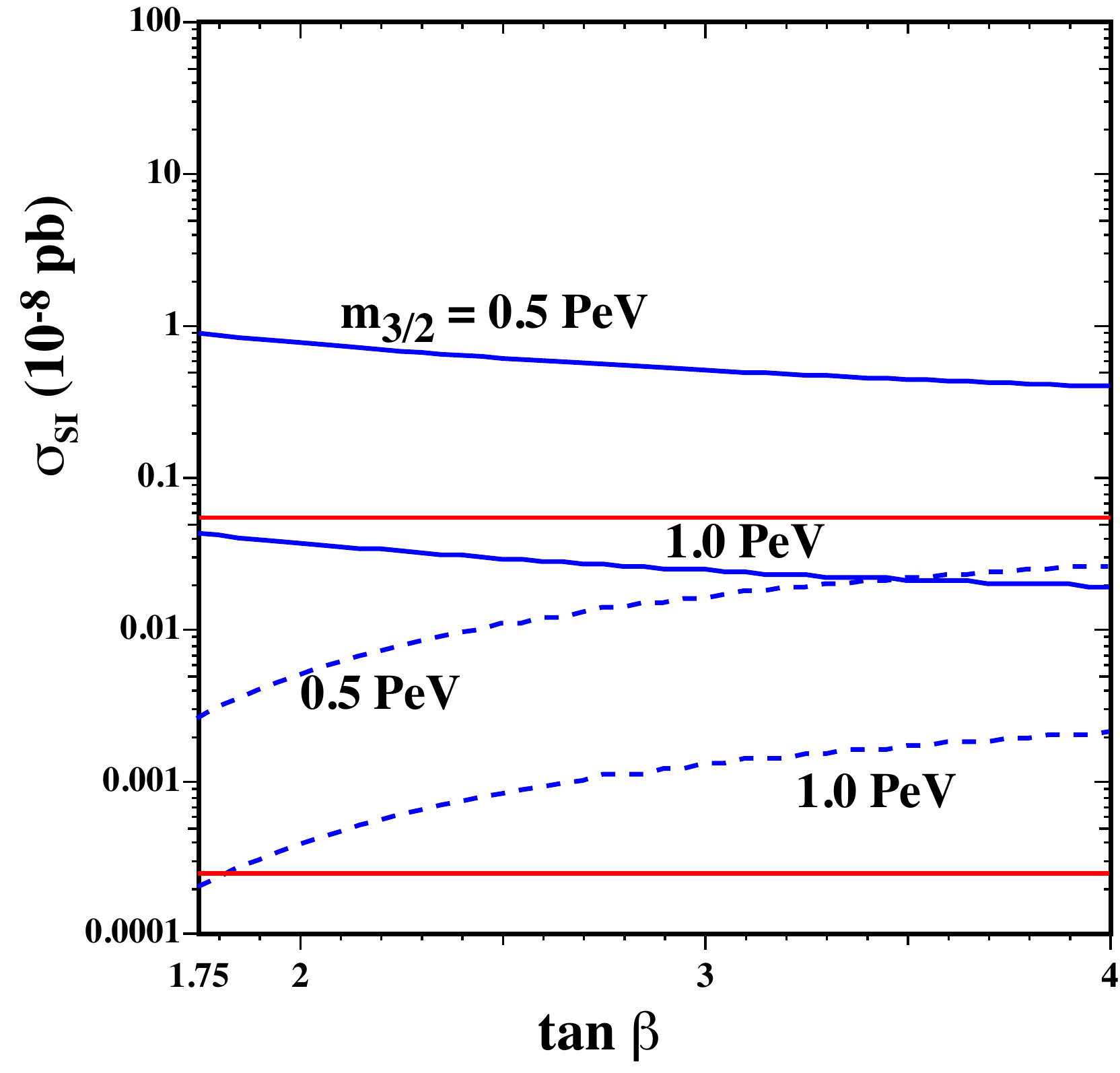}
\caption{\it The spin-independent elastic cross section for LSP scattering on protons as a function of $m_{3/2}$ (left) for $\tan \beta = 1.8$, 2.0, and 2.2, and as a function of $\tbt$ (right) for $m_{3/2} = 0.5$ and 1.0 PeV. In both panels, $\mu = 1080$ GeV (solid) and -900 GeV (dashed). The upper and lower horizontal lines in the left panels correspond to the limit from PandaX-4T \cite{PandaX} and the neutrino floor respectively.   }
\label{fig:1DSI}
\end{figure}

Similarly, we plot in Fig.~\ref{fig:planeSD}, the spin-dependent cross sections, $\sigma^{SD}_p = \sigma_2$, for $\tbt = 1.8$ (left)
and 2.2 (right). Cross section values are again as labelled in units of $10^{-8}$ pb. The current best experimental limit on the spin-dependent cross section comes from 
PICO \cite{PICO} and is too weak to be visible on this plane. The strongest constraint $2.5 \times 10^{-5}$ pb for 
$m_\chi = 25$ GeV, and is $3.5 \times 10^{-4}$ for $m_\chi = 1$ TeV.  Once again we find a rapid variation in the cross section in the region transitioning between a wino
and Higgsino LSP. This is also seen in the 1-dimensional plots shown in the left panel of Fig.~\ref{fig:1DSD}. As one can see, there is far less dependence on the sign of $\mu$ for the spin-dependent cross section. This can be readily understood by comparing 
Eqs.~(\ref{eq:al2App}) and (\ref{eq:al3App}). 

\begin{figure}[!ht]
\centering
\includegraphics[width=8cm]{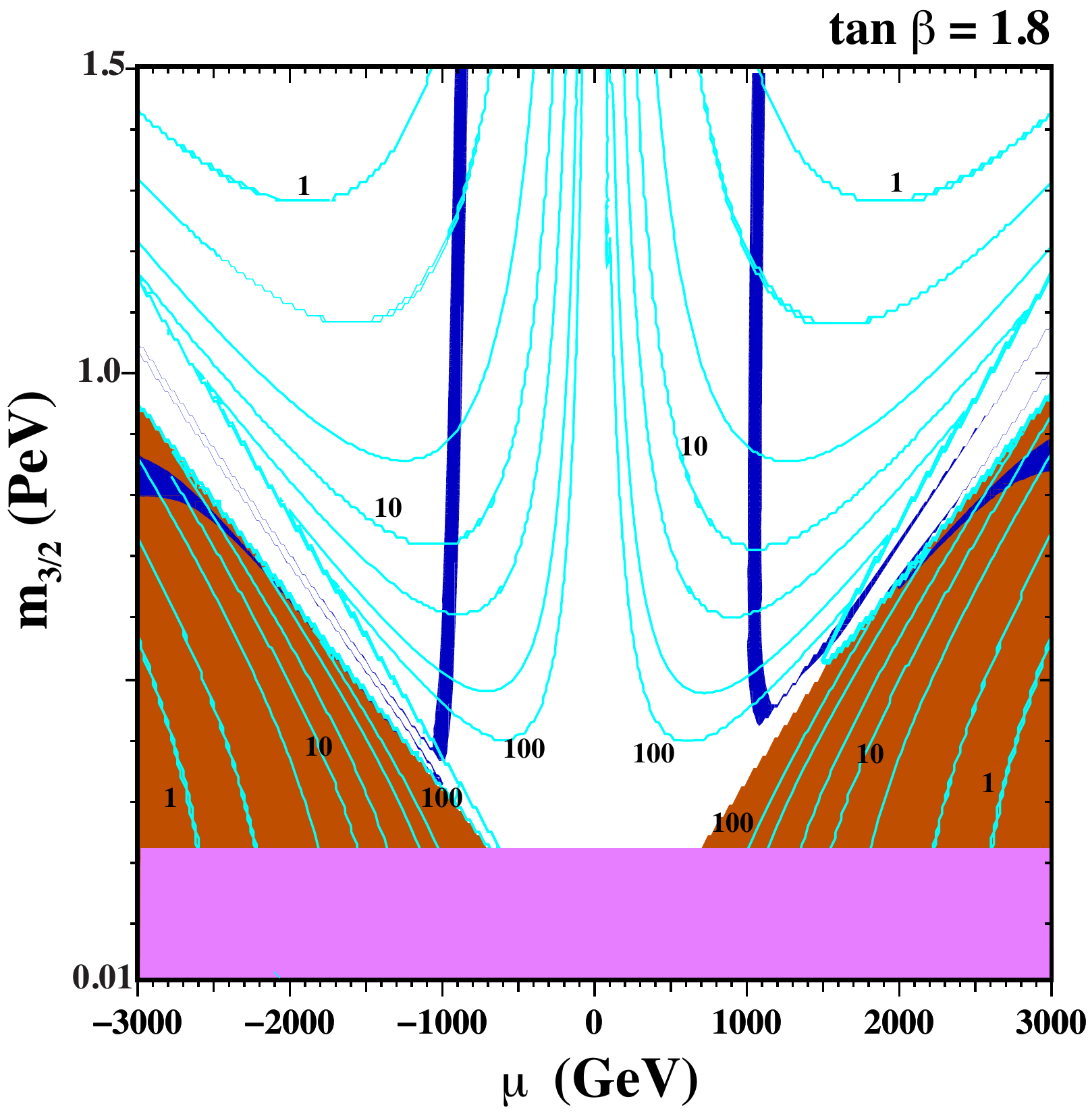}
\includegraphics[width=8cm]{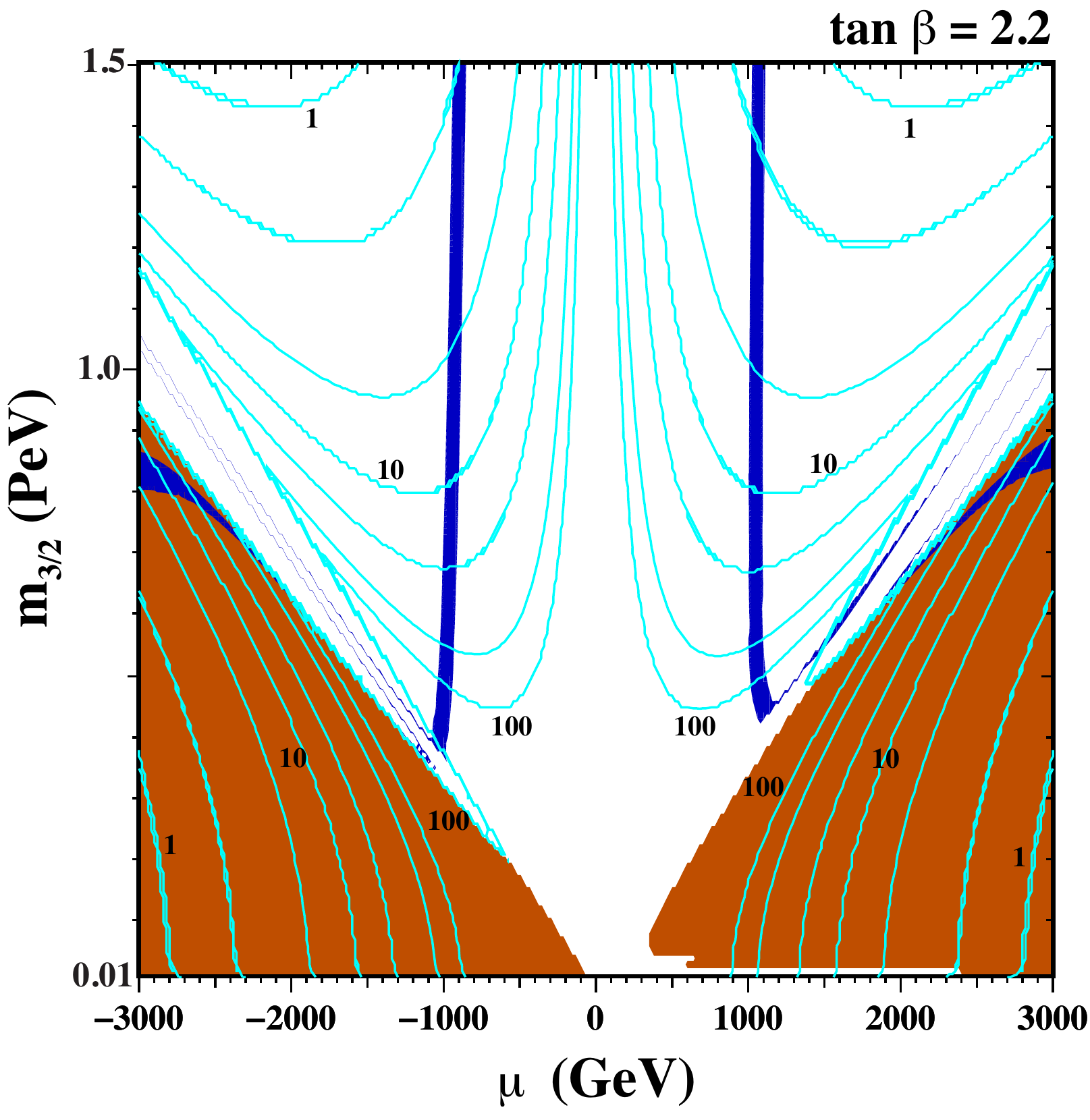}
\caption{\it The spin-dependent elastic cross section for LSP scattering on protons in the ($\mu, m_{3/2}$) plane for fixed $\tbt = 1.8$ (left) and 2.2 (right). The shaded regions are the same as in Fig.~\ref{oldplanes}.  Values of the cross sections are as labelled in units of $10^{-8}$ pb. Contours between the labeled decades are 2 and 5 $\times$ the preceding decade. }
\label{fig:planeSD}
\end{figure}

\begin{figure}[!ht]
\centering
\includegraphics[width=8cm]{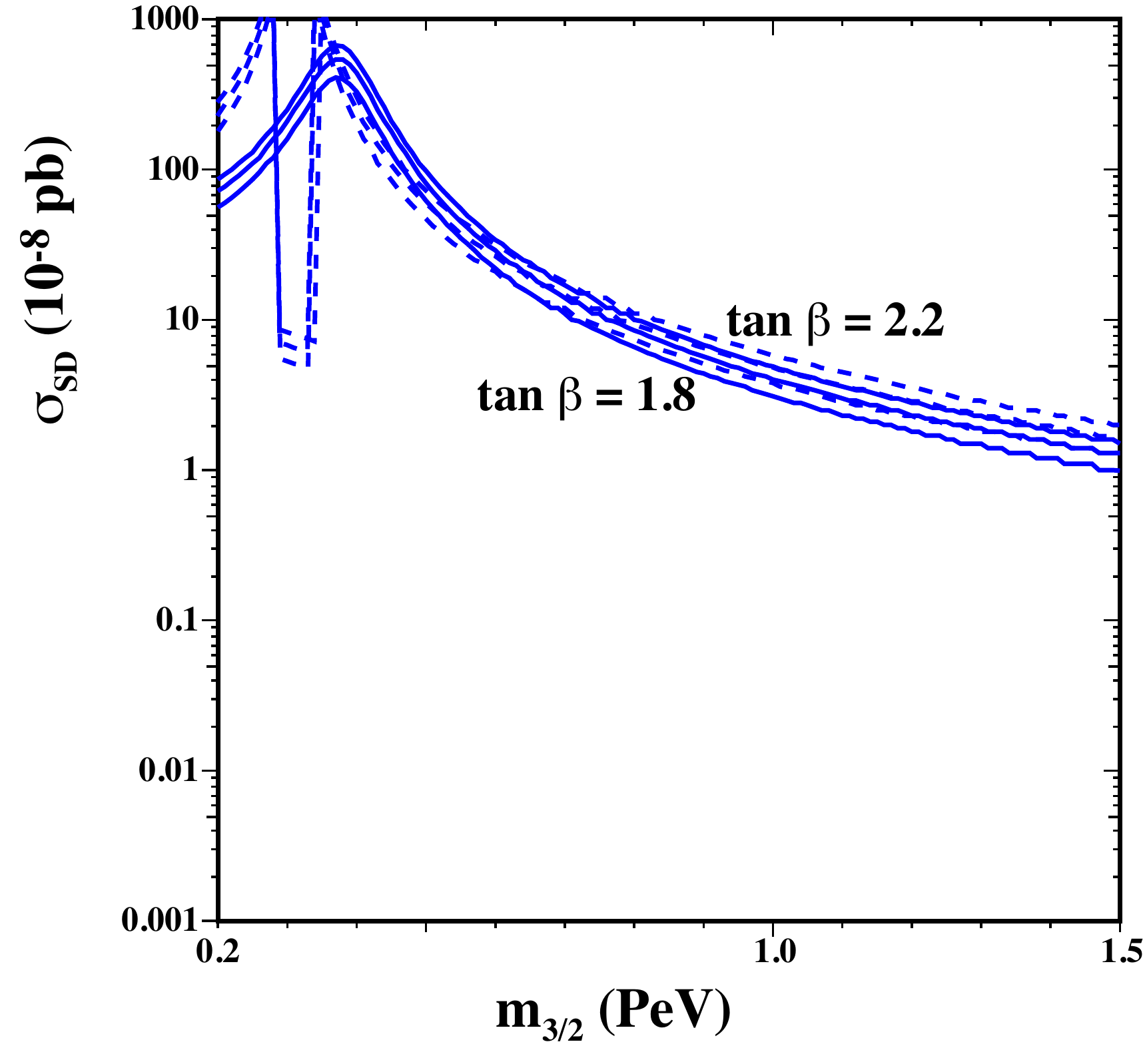}
\includegraphics[width=8cm]{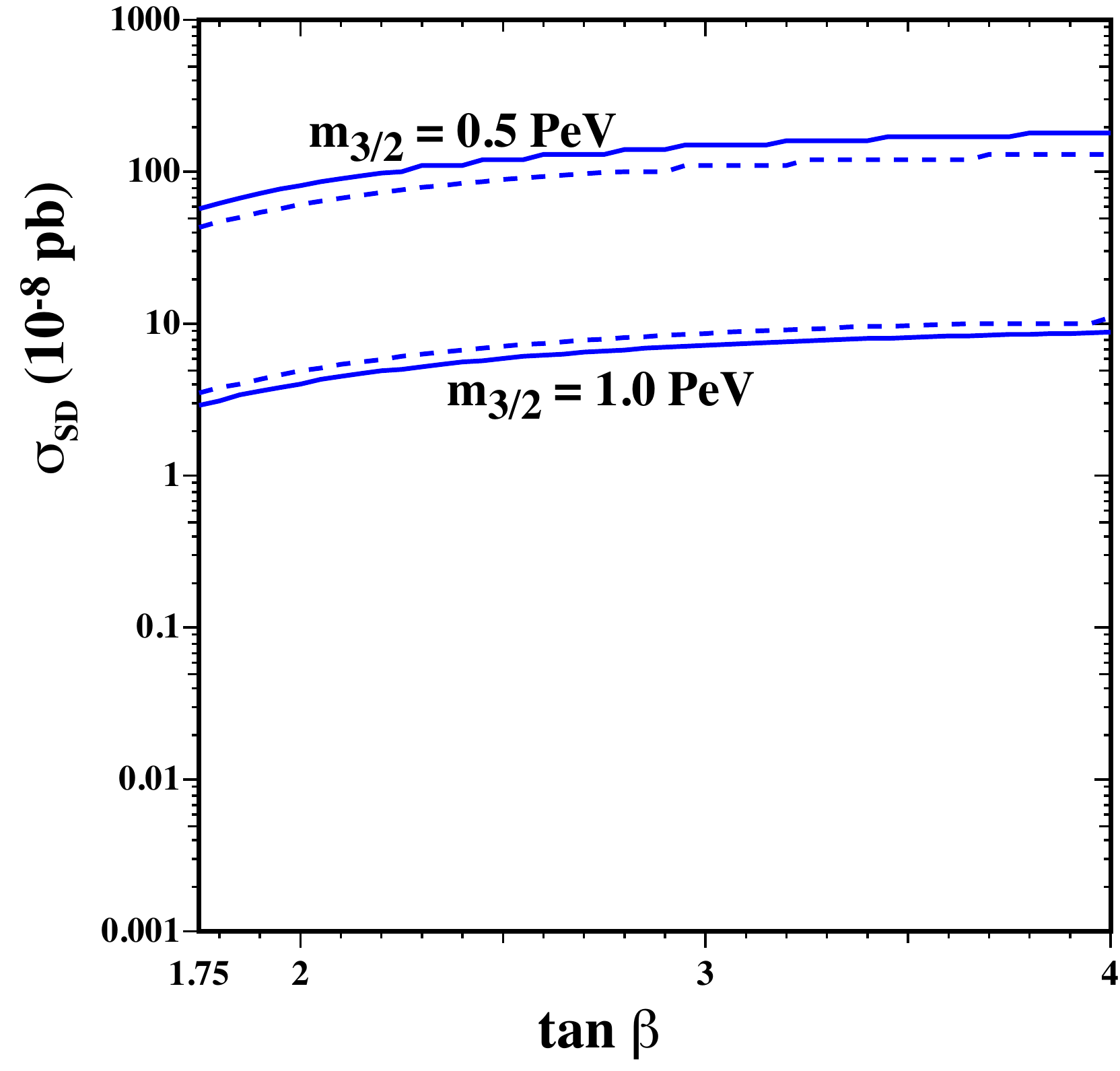}
\caption{\it The spin-dependent elastic cross section for LSP scattering on protons as a function of $m_{3/2}$ (left) for $\tan \beta = 1.8$, 2.0, and 2.2, and as a function of $\tbt$ (right) for $m_{3/2} = 0.5$ and 1.0 PeV. In both panels, $\mu = 1080$ GeV (solid) and -900 GeV (dashed). }
\label{fig:1DSD}
\end{figure}

Finally, we present in Fig.~\ref{fig:planeT} results in the $(m_T/m_{3/2}, m_{3/2})$ plane where we allow the stop masses
to be non-universal. More specifically,
we set the boundary condition for the 3rd generation 
left-handed quark doublet and right-handed stop to be $m_T$ at the high energy supersymmetry breaking input scale.  We again display results for $\tbt = 1.8$ (upper panels) and 2.2 (lower panels) and two fixed values of $\mu = -900$ GeV (left) and 1080 GeV (right).
Because we fix $\mu$ at a value which leads to $\Omega_\chi h^2 \simeq 0.12$ in the models with universal sfermion masses, large portion of these planes provide the correct relic density as seen by the dark blue shading. For small values of $m_{3/2}$, the LSP is predominantly a wino in the dark red shaded region.  For 
small values of $m_T/m_{3/2}$, one of the stops becomes tachyonic and we shade this region pink. The Higgs masses are shown by red dot-dashed curves as labelled.  
As one can see in much of the dark blue shaded region with good relic density, the Higgs mass similar to the experimental value within uncertainties. Furthermore, for $\tbt = 2.2$, more of the parameter space is in agreement with experiment compared to the case of $m_T=m_{3/2}$ as the Higgs mass decreases with $m_T/m_{3/2}$. This means larger values of $\tbt$ will have some viable parameter space for $m_T/m_{3/2}$ small. However, it is clear from these figures that $\tbt$ can still not be significantly larger than $2.2$

\begin{figure}[!ht]
\centering
\includegraphics[width=7.8cm]{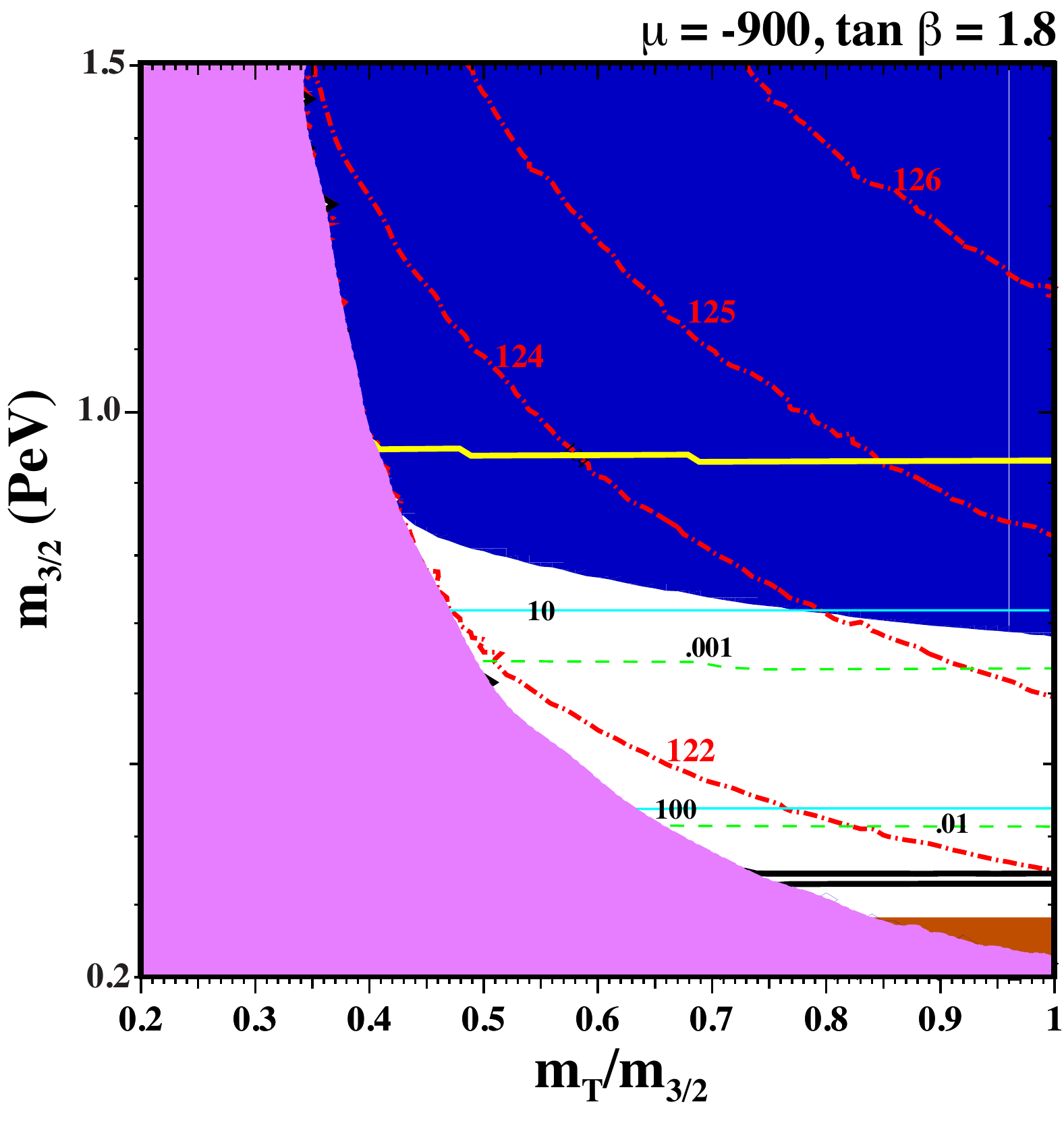}
\includegraphics[width=7.8cm]{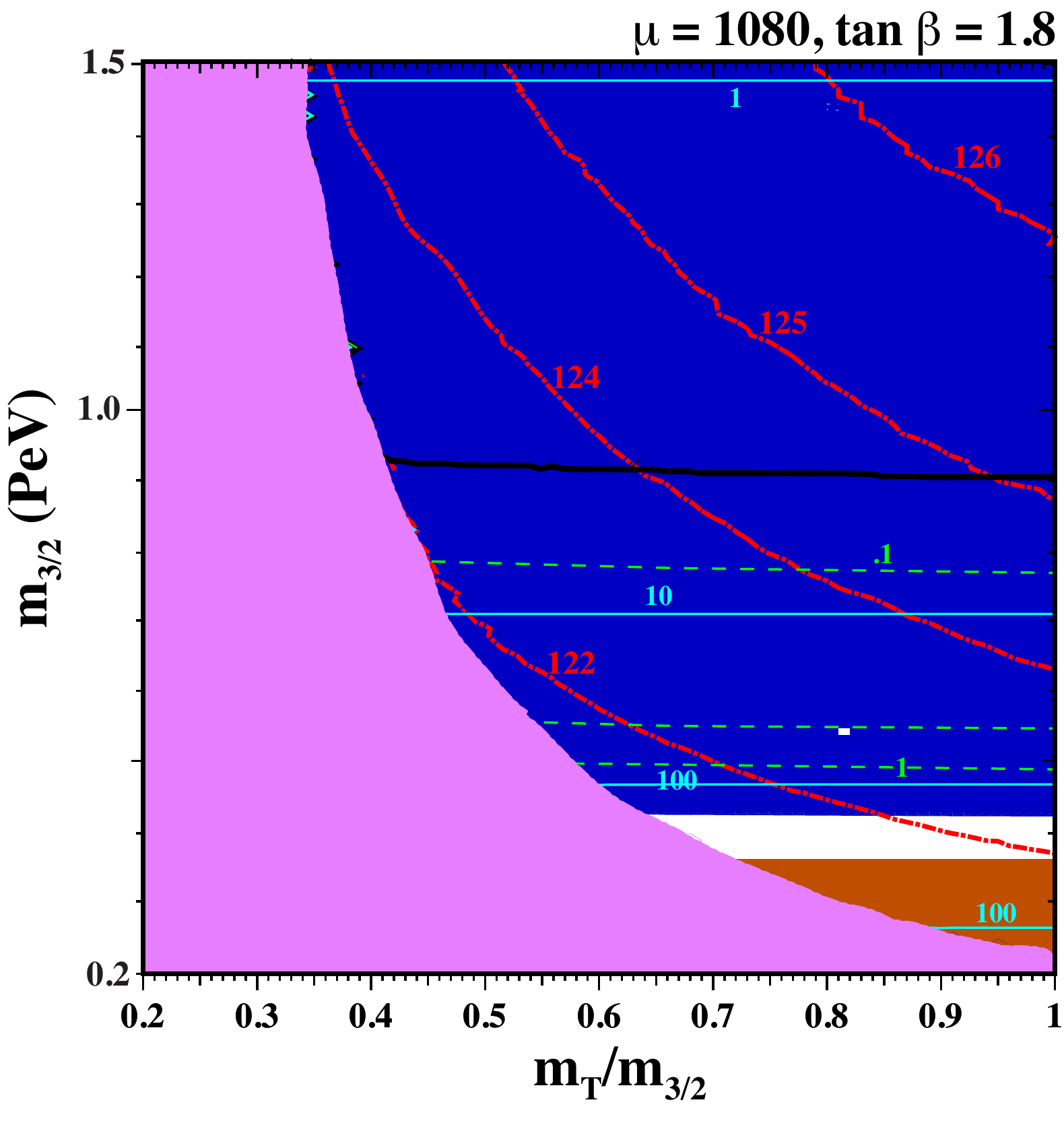}\\
\includegraphics[width=7.8cm]{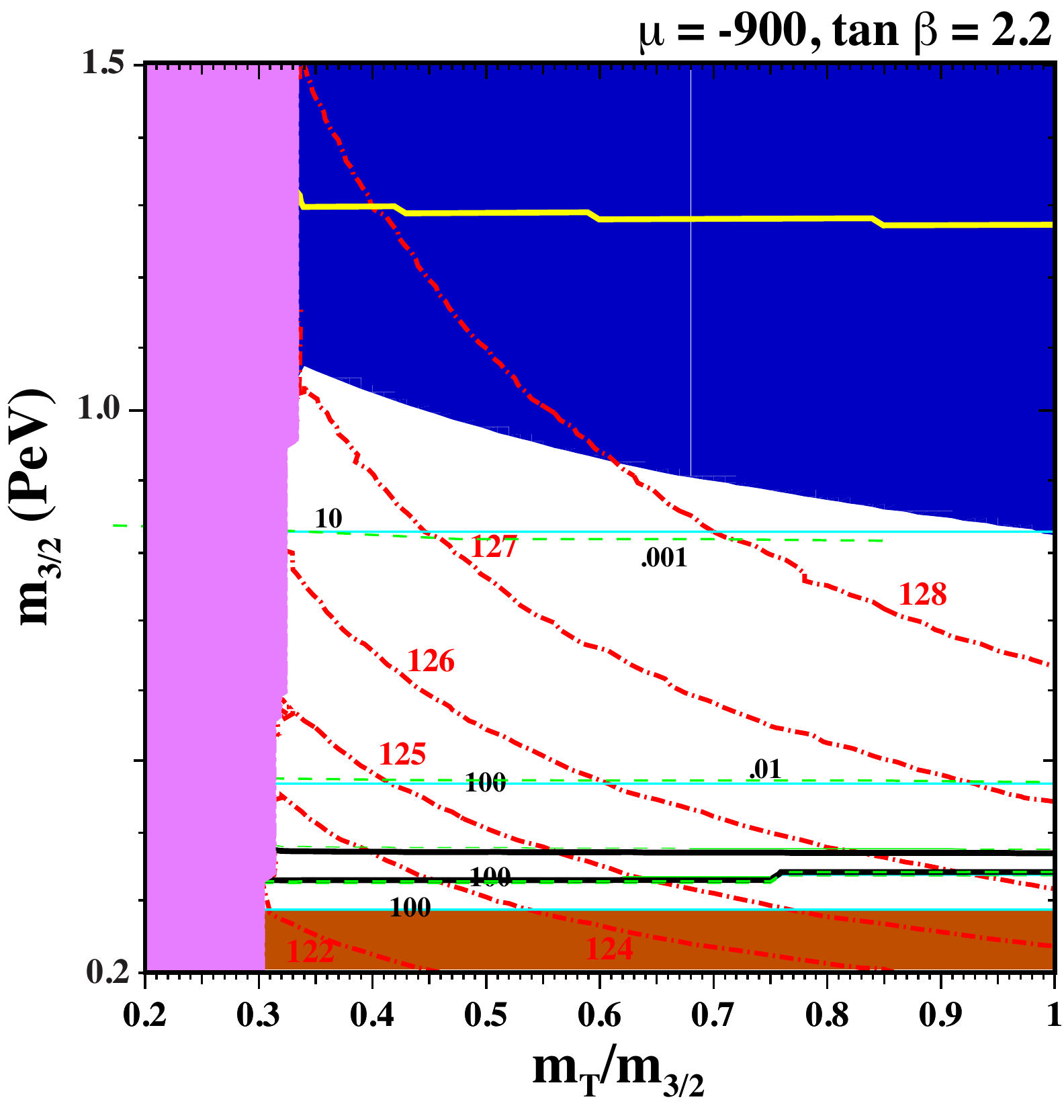}
\includegraphics[width=7.8cm]{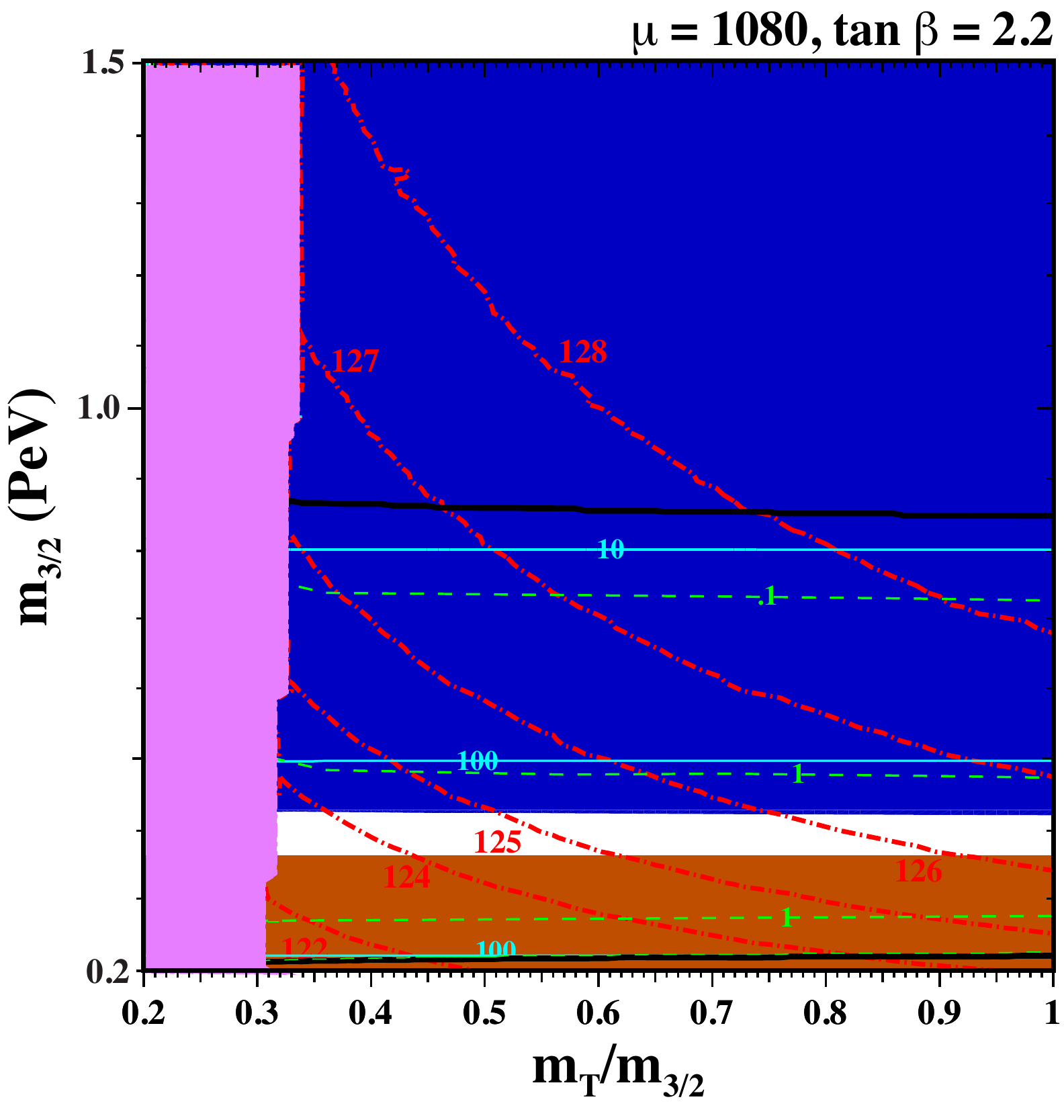}
\caption{\it The $(m_T/m_{3/2}, m_{3/2})$ plane for fixed $\tan \beta = 1.8$ (upper) and 2.2 (lower) for
fixed values of $\mu = -900$ GeV (left) and 1080 GeV (right). The pink shaded region is excluded
as it contains a tachyonic stop. In the dark red shaded region, there is a wino LSP. In the remainder of the plane, the Higgsino is the LSP. Higgs mass contours, with masses labelled are shown as red dot-dashed curves. Spin-independent cross sections are shown by green dashed lines and spin-dependent cross sections by solid blue lines. For spin-independent cross sections, the experimental upper limit  is shown by the solid black line and the neutrino floor by a yellow line.  }
\label{fig:planeT}
\end{figure}

The elastic scattering cross sections are relatively insensitive to the stop masses and are shown in Fig.~\ref{fig:planeT} as nearly horizontal lines. Spin-independent cross sections are shown by green dashed lines and spin-dependent cross sections by solid blue lines labelled in units of $10^{-8}$ pb. Once again, we see the experimental upper limit (solid black line) and the neutrino floor (yellow line) for $\mu < 0$, applicable for the spin-independent cross section only. Points below the black line are in excess of the experimental bound and points above the yellow line fall beneath the neutrino floor.

\section{Conclusions}

Experimental verification of physics beyond the Standard Model is of the utmost importance. We know such physics must be present in order to account for dark matter as well as other aspects of cosmology
such as the baryon asymmetry. Supersymmetry 
is a well studied extension which
helps better explain features of the Standard Model as well as providing a dark matter candidate. However because the mechanism for breaking supersymmetry is unknown, there is 
a very diverse set of supersymmetric models to study.  Models such as mSUGRA with weak scale supersymmetry breaking are under considerable pressure, as weak scale superpartners have yet to be discovered \cite{nosusy}. Some high-scale models with an
EeV scale gravitino and still higher superpartner masses are extremely challenging
from the point of view of discovery \cite{EeV}. Naively, one might think that models with PeV scalar masses would present similar challenges.

We have considered here, PGM models with a
PeV gravitino mass (and similarly massive scalars). However, the gaugino masses in these models are loop suppressed and may be of order of $\sim 1$ TeV. Further, we have extended the minimal model (with two parameters - $m_{3/2}$ and $\tbt$) to
include $\mu$ as a free parameter, thus
easily allowing for the possibility of Higgsino dark matter. Despite the high scalar masses, which allows us to consider the decoupling limit, the dark matter-proton
scattering cross sections are {\em not} unobservably small. We considered the 
$\mu, m_{3/2}$ parameter space for fixed $\tbt$, taking values of $|\mu| \le 3$ TeV, 
and $m_{3/2} \le 1.5$ PeV with $\tbt = 1.8 $ and 2.2. For higher values of $\tbt$, the calculated Higgs mass becomes significantly larger than it measured value (even taking into account theoretical uncertainties in the calculation). To increase the parameter space, we also allowed for the possibility that the stop masses are not universal at the input supersymmetry breaking scale (which we have taken to be the GUT scale). This effect, however was marginal. 

We found that the spin-independent cross section was quite sensitive to the sign of $\mu$. For $\mu > 0$, the current experimental constraint on $\sigma^{SI}_p$
from PandaX-4T \cite{PandaX} excludes gravitino masses below 850 TeV when $\mu \sim 1$ TeV ($m_{3/2} \lesssim 200$ TeV are allowed, as the cross section begins to drop 
when the LSP is wino-like). At larger $m_{3/2}$, the elastic cross sections are large enough to be detected in future experiments. For example, at $m_{3/2} = 1.5$ PeV (with $\mu \sim 1$ TeV), $\sigma^{SI}_p \approx 10^{-10}$ pb well above the neutrino 
floor at $\approx 2.5 \times 10^{-12}$ pb for $m_\chi \simeq 1$ TeV. In contrast, for $\mu < 0$, the cross section is below the neutrino floor when $m_{3/2} \gtrsim$ 1 PeV and is somewhat sensitive to $\tbt$ (see Fig.~\ref{fig:1DSI}). Current experimental results do not place significant bounds on the parameter space when $\mu < 0$. The resulting cross section for spin-dependent interactions remain at least two orders of magnitude below current experimental bounds.
Nevertheless, we remain hopeful that a signal
for Higgsino dark matter in PGM-like models is viable in future direct detection experiments.

\section*{Acknowledgements}
J.L.E. would like to thanks Tsutomu T. Yanagida for useful discussions during the early stages of this work. The work of K.A.O.~was supported in part by DOE grant DE-SC0011842  at the University of
Minnesota.

\end{document}